\documentclass[preprint]{aastex}

\slugcomment{Paper submitted to Astronomical Journal on 2004 Sep 27}
\shorttitle{Exploring Halo Substructure. VI.}
\shortauthors{Majewski et al.}

\begin{document}
\title{Exploring Halo Substructure with Giant Stars. VI.  Extended
  Distributions of Giant Stars Around the Carina Dwarf Spheroidal Galaxy
  --- How Reliable Are They? (\small astro-ph/0503627) }

\author{Steven R. Majewski\altaffilmark{1,2},
  Peter M. Frinchaboy\altaffilmark{1,2},
  William E.  Kunkel\altaffilmark{2,3}, 
  Robert Link\altaffilmark{1,4}, 
  Ricardo R. Mu\~noz\altaffilmark{1},
  James C. Ostheimer\altaffilmark{1,5}, 
  Christopher Palma\altaffilmark{1,6}, 
  Richard J. Patterson\altaffilmark{1}, 
 and Doug Geisler\altaffilmark{7}}

\altaffiltext{1}{Department of Astronomy, University of Virginia,
  Charlottesville, VA 22903-0818; 
  srm4n@virginia.edu, 
  pmf8b@virginia.edu,
  rrm8f@virginia.edu,
  jco9w@alumni.virginia.edu,
  robert.link@ngc.com, 
  ricky@virginia.edu.}

\altaffiltext{2}{Visiting Astronomer, Cerro Tololo Inter-American
  Observatory, National Optical Astronomy Observatory, which is operated
  by the Association of Universities for Research in Astronomy, Inc.,
  under cooperative agreement with the National Science Foundation. }

\altaffiltext{3}{Las Campanas Observatory, Carnegie Institution of
  Washington, Casilla 601, La Serena, Chile; kunkel@jeito.lco.cl.}

\altaffiltext{4}{Present address: Northrop Grumman Information
  Technology - TASC, 4801 Stonecroft Blvd., Chantilly, VA 20151.}

\altaffiltext{5}{Present address: 1810 Kalorama Rd., NW, \#A3,
  Washington, D.C. 20009.}

\altaffiltext{6}{Department of Astronomy \& Astrophysics, Pennsylvania
  State University, University Park, PA 16802; cpalma@astro.psu.edu.}

\altaffiltext{7}{Department of Physics, Universidad de Concepci\'{o}n,
  Concepci\'{o}n, Chile; dgeisler@astro-udec.cl}

\begin{abstract}
  
  The question of the existence of active and prominent tidal disruption
  around various Galactic dwarf spheroidal (dSph) galaxies remains
  controversial.  That debate often centers on the nature (bound versus
  unbound) of extended populations of stars claimed to lie outside the
  bounds of single King profiles fitted to the density distributions of
  dSph centers.  However, the more fundamental issue of the {\it very
    existence} of the previously reported extended populations is still
  contentious.  We present a critical evaluation of the debate centering
  on one particular dSph, Carina, for which claims both for and against
  the existence of stars beyond the King limiting radius have been made.
  Our review includes a detailed examination of all previous studies
  bearing on the Carina radial profile and shows that among the previous
  survey methods used to study Carina, that which achieves the highest
  detected dSph signal-to-background in the diffuse, outer parts of the
  galaxy is the Washington $M,T_2$ + $DDO51$ filter approach from Paper
  II in this series, which depends on the stellar surface gravity
  sensitivity of the $DDO51$ filter to remove the bulk of contaminating
  foreground stars and leave predominantly stars on the Carina red giant
  branch.  The second part of the paper addresses statistical methods
  used to evaluate the reliability of $M,T_2,DDO51$ surveys in the
  presence of photometric errors and for which a new, \textit{a
    posteriori} statistical analysis methodology is provided.  This
  analysis demonstrates that the expected level of contamination due to
  photometric error among the photometrically-selected candidate Carina
  giant sample stars in Paper II is no more than 13--27\% --- i.e., only
  slightly higher than originally predicted in Paper II.  In the third
  part of this paper, these statistical methods are tested by new Blanco
  + Hydra multifiber spectroscopy of stars in the $M,T_2,DDO51$-selected
  Carina candidate sample.  The results of both the new \textit{a
    posteriori} and the previous Paper II contamination predictions are
  generally borne out by the spectroscopy: Of 74 candidate giants with
  follow-up spectroscopy, the $M,T_2,DDO51$ technique successfully
  identified 61 new Carina members, including 8 stars outside the
  photometrically defined King profile limiting radius.  In addition,
  among a sample of 29 stars that were {\it not} initially identified as
  candidate Carina giants but that lie just outside of our selection
  criteria, 12 have radial velocities consistent with membership in
  Carina, including 5 extratidal stars.  The latter shows that, if
  anything, the Paper II estimates of the Carina giant star density
  outside the King limiting radius may have been underestimated.  Carina
  is shown to have an extended population of giant stars extending to a
  major axis radius of $40\arcmin$, i.e., 1.44 times the nominal King
  limiting radius.  A number of bright blue stars are also found to have
  the radial velocity of Carina and we discuss the possibility that some
  of them may be part of the Carina post-asymptotic giant branch
  population.  A few additional radial velocity members are found to lie
  among stars in the horizontal branch and anomalous Cepheid regions of
  the Carina color-magnitude diagram.

\end{abstract}

\keywords{Galaxy: evolution -- Galaxy: halo --
Galaxy: structure -- stars: galaxies: individual (Carina dSph) --
photometry -- stars: giants -- methods: statistical }

\section{Introduction}

\subsection{The Search for Extratidal Stars around dSph Satellites}

The question of the stability and response of satellite dwarf spheroidal
(dSph) galaxies to the Milky Way tidal field has remained a central
issue in the studies of these systems since Hodge's (\citeyear{hod64})
early discussion of tides in the Ursa Minor system.  The question bears
not only on the properties of these satellites --- e.g., their mass,
dark matter content, shape, internal dynamics, substructure, etc. ---
but also potentially on their role in supplying material for the
continuing growth of their parent systems.  Finding populations of dSph
stars now unbound from their parent would certainly provide definitive
evidence of ongoing stellar mass loss --- presumably through tidal
stripping --- while measuring the rate and location of stars leaving the
dSph are important gauges of the mass ratio of the dSph to the Milky Way
(e.g., \citealt{hod64}; more recently,
\citealt*{moo96,bur97,JSH99,JCG02}) as well as the stellar accretion
rate of the Galaxy.  However, with the exception of the impressive and
now well-studied Sagittarius dSph system, measurements of such
``extratidal'' populations emanating from nearby dSph galaxies has
proven difficult to undertake and the results of such studies difficult
to interpret.  The outlying regions where the putative stripping occurs
are at extremely low surface brightness levels (typically $\Sigma_{\rm V} >31$
mag arcsec$^{-2}$), and the mere detection of extratidal debris against
a significant Milky Way foreground remains a daunting technical
prospect.

Nevertheless, extratidal RR Lyrae stars were first reported around the
Sculptor dSph by \citet{vA78}, and the suggestion that there were
extratidal Sculptor stars was supported by studies of other Sculptor
tracer stars by \citet{esk88a}, \citet*[][``IH95'']{IH95},
\citet{wes00}, \citet*[][``W03'']{W03} and \citet{wes05}.  A review of
the literature finds at least one report of potential ``extratidal
stars'' around every Milky Way dSph: Sextans \citep{gou92}, Fornax
\citep{esk88b}, Draco \citep*{Smith97,KK00}, Leo II \citep*{SMP00}, Leo I
\citep{soh03}, Ursa Minor \citep{md01,pal03}, Sagittarius \citep[see
summary of sources in Fig.\ 17 of][]{sgr1}, and, of course, around the
Carina dSph (see \S1.2).  In their comprehensive, systematic,
photographic starcount study of Milky Way dSphs, \citeauthor{IH95} found
suggestions of excesses of stars beyond the nominal limiting radius of
almost every Galactic dSph (and suggested that Sextans and Sculptor were
the best candidates for tidally disrupted systems among the dwarfs they
studied).

Beyond the difficulties of mere detection, determining just exactly what
any discovered excesses of ``extratidal'' stars {\it are} also remains a
challenging problem.  Often the term ``extratidal'' is invoked to mean
stars beyond the King `tidal' radius of one-component models fitted to
the density profile of the system; the limiting radii of fitted King
functions are sometimes attributed to true tidal boundaries because of
the similarity in appearance of at least some dSph radial profiles to
that of tidally-truncated systems.  However, a number of recent surveys
(see references in previous paragraph) have found beyond the King
profile-like central components of dSphs the presence of additional,
extended components having a more gradual (e.g., power law) radial
density decline.  Such two-component density profiles are generated
naturally in N-body models of the tidal disruption of satellite galaxies
where the second population, outside of where the density profile
``breaks'', corresponds to unbound tidal debris particles
\citep[e.g.,][]{JSH99, may02}.  However, although tidally-truncated,
King-like profiles have remained useful descriptors of the inner
structures of dSph galaxies for quite some time
\citep[e.g.,][]{hod61a,hod61b}, there is no good reason to be fitting
actual King (\citeyear{kin62,kin66}) functions to systems with such long
crossing times, and in some cases more gradually declining (e.g.,
exponential) functional forms than King functions have been reported to
yield equally suitable fits to the {\it entire} density profiles of
Galactic dSphs (\citealt{FL83}, \citeauthor{IH95}, \citealt*{ACM01,
  ode01b, pal03}, \citeauthor{W03}).  Even in systems for which King
functions {\it are} appropriate descriptors, e.g., globular clusters,
the King limiting radius may only be an approximation to the true,
instantaneous Roche limit \citep[e.g.,][]{kin62,JCG02,hay03}, which, for
example, naturally varies in position at different phases of eccentric
orbits (i.e., as a function of position in the Galactic potential).
Thus, even in the simplest interpretations of the structure of dSph
galaxies the actual state --- bound or unbound --- of stars beyond the
King limiting radius is not clear {\it a priori}, and any use of the
expression ``extratidal'' in this paper is meant only to signify that
stars lie beyond the single fitted King limiting radius, {\it not} that
the stars are unbound.\footnote{We will also use the more general
  expression ``break population'' to refer to those stars inhabiting the
  gradually declining density law that ``breaks'' from a King profile at
  large radii, creating an ``excess'' population of stars not accounted
  for by a single King profile.}
  
Moreover, it is not clear whether any dark matter in these dSph systems
actually follows the light profile.  Indeed, the possibility of
multi-component structures within dSph systems is central to the debate
over the meaning of ``extratidal excesses'' as well as the mass-to-light
ratios of these systems.  The possibility that the ``extratidal''
excesses of stars are only a second population of dSph stars that
actually lie bound deeply within large dark matter halos (like those of
\citealt{sto02}) has also been postulated \citep{bur97, hay03}.
Measuring the dynamics of stars in these ``extratidal'' populations
should help resolve whether these excess stars are bound or unbound and
whether mass follows light in dSphs \citep{kro97,kle99}, but doing this
has proven extremely challenging, and few examples of dSph stars clearly
outside the King limiting radius of their parent system having measured
radial velocities yet exist, apart from those stars observed in the
tails of the the Sagittarius system --- for which the first substantial
results have only been recently obtained \citep*{Y03,sgr2,vivas05}
Although in this paper we present (\S3) the first significant sample of
spectroscopic observations of stars beyond the King radius of any other
dSph, and although some of these data are of sufficient level of
accuracy and degree of separation from the dSph core to weigh in on the
issue of whether the photometrically discovered Carina ``excess'' stars
are bound or not, we leave this particular aspect for a separate
discussion (supplemented with additional data; R. Mu\~noz et al., in
preparation).  Our primary concern here is not to {\it explain} the
Carina break population, but, because this point alone has been
contentious, to {\it prove that it is real}.

\subsection{The Carina dSph}

Although the question of the extended structure of all of the dSphs
(apart from Sagittarius) has remained controversial, the case of the
Carina dSph has received particular attention over the past decade.  A
critical review of previous work on this dSph would seem in order given
that the earlier three claims for a detected break population have been
apparently refuted by two more recent studies.  Such a review is the
first goal of this paper.  In the remainder of this section we summarize
and compare the previous photometric studies of Carina, with a
particular emphasis on assessment of background levels, the primary
limitation to reliable detection of diffuse features like dSph break
populations and tidal tails.  In \S2 we focus on the question of
modeling the contamination rate of photometrically-selected dSph giant
star surveys, and present a new analytical method for assessing this
rate within our previous survey of Carina in Paper II.  Finally (\S3) we
test the predictions of the three extant contamination rate analyses of
the Paper II results using new spectroscopic data on a subsample of
stars in the Carina field.  We also report the presence of some curious
blue stars in the field having Carina velocities.

\subsubsection{Starcount Studies of the Carina dSph}

\citeauthor{IH95} was the first report of an excess beyond the King
limiting radius of the 101 kpc-distant (Mateo 1998) Carina system (see
data in Fig.\ 1).  Using photographic starcounts to $B\sim22$
($R\sim21$) in a survey encompassing $3\arcdeg \times 3\arcdeg$ areas
centered on each of the Galactic dSphs, \citeauthor{IH95} found at large
radii the presence of excess stars with respect to King profiles in most
cases.  The \citeauthor{IH95} discussion of this phenomenon includes
consideration of the possibility that these excess stars represent tidal
stripping from the parent dSph systems.

Another, more recent dSph study that employed similar starcounting
techniques, but with deeper (albeit single filter) CCD data, is that of
\citeauthor{W03}.  We include their density profile and fitted King
profile in Figure 1.  Note that \citeauthor{W03}'s King profile is
similar to that found by \citeauthor{IH95}, with a King limiting radius
only slightly larger ($31\farcm8$) than the $28\farcm8$ radius of
\citeauthor{IH95}.  However, while claiming to find possible extratidal
debris in their similar study of the Sculptor dSph, \citeauthor{W03}
actually ``rule out'' the existence of a Carina break population at the
levels claimed by both \citeauthor{IH95} and by \citet*[][``Paper II''
hereafter]{paperII}.  As \citeauthor{W03} state, the discrepancy between
their results and the \citeauthor{IH95}/\citeauthor{paperII} surveys is
``disturbing'' and ``difficult to explain''.  Lacking more details about
each survey, we cannot offer a definitive explanation for the
differences in findings, but do offer several pertinent observations
that highlight possibilities.

As stressed by \citeauthor{IH95} (for example) a major challenge
confronting the study of the low surface brightness outskirts of dSphs
is a proper accounting of the background.  Background overestimation can
erase faint dSph features, whereas background underestimation can
inflate or artificially produce the appearance of diffuse features in
the outer parts of dSphs.  In the case of the two starcount studies
mentioned, the measured background levels are comparable to the density
of stars from the dSph {\it inside} the King limiting radius (Fig.\ 1).
The actual ``extratidal'' signal that has been reported for Carina by
any survey is significantly {\it smaller} than the background that is
being subtracted by either \citeauthor{IH95} or \citeauthor{W03}.  The
background confronted by \citeauthor{IH95} has actually been
significantly exacerbated by background galaxies, which were not
excluded from the source counts (because morphological discrimination
became unreliable in their data at the magnitude limit that IH95 adopted
to access large numbers of Carina stars).  Although they do use
morphological criteria to exclude galaxies in their survey,
\citeauthor{W03} also note that stars and galaxies are ``almost
indistinguishable'' at the faint end of their survey; nevertheless,
morphology was used to remove some 4-10\% of the total sources from
their Carina survey.  Yet galaxy contamination can be an order of
magnitude or more {\it larger} than that expected from Milky Way stars
at the brightness limits of either of these studies \citep[see, for
example, Fig.\ 1 of][]{RM93}.  A clue that extragalactic sources may
play a dominant (but unevaluated) role in the background considerations
of these surveys is that although the \citeauthor{W03} survey probes
several magnitudes {\it fainter} than that of \citeauthor{IH95}, the
reported background density of sources is some four times {\it lower} in
the former than in the latter survey.

The signal-to-background ratio in the outer dSph density profile is
critically sensitive to both potential random and systematic errors and
drives sensitivity to diffuse structures: (1) One expects naively that a
study with a higher mean background compared to the central dSph density
should be commensurately less sensitive to diffuse features, which can
become ``lost'' in the Poissonian fluctuations of the background.  (2)
Since it is typical to take as the ``background level'' the point in the
dSph radial profile where it flattens out, a study with a higher mean
background to dSph density will be less sensitive to where the profile
``goes to zero'', will tend to underestimate where this happens, and
will include the lost, tenuous signal of outer dSph stars as background.
Both effects are especially serious in studies where the background is
larger than the ``extratidal'' signal of interest, but mitigated by
increased survey area: (1) The {\it error in determination of the mean
  background}, which plays a critical role in the overall reliability of
the background-subtracted density profile, correlates with the inverse
square root of the number of background stars used to evaluate the
background density.  A larger survey area provides proportionately
larger numbers of background sources at the same magnitude limit.  (2) A
larger survey area provides an overall larger sampling of sky away from
the dSph itself, thus diluting the effects of ``contamination'' of the
background by dSph stars.

In these respects, it is interesting to note that \citeauthor{IH95},
which has a background level about 1/4 their measured central Carina
density, {\it does} detect an extratidal excess beginning at about 1/15
the central density whereas \citeauthor{W03} do not see this excess in
their survey, which adopts a background at nearly 1/40 their central
Carina density.  The difference in findings may relate to the {\it error
  in the mean} of the derived backgrounds.  From their analysis of the
periphery of a 9 deg$^2$ survey field, \citeauthor{IH95} claim a rather
small fractional error in their background evaluation --- 0.7\%.
\citeauthor{W03} do not give an estimated error in their determined
background level, but this would need to have been determined to better
than about 3\% in order to have the same absolute error as
\citeauthor{IH95} (0.02 background sources arcmin$^{-2}$), whereas the
amount of assumed ``dSph-free'' area is smaller in their 4 deg$^2$
survey compared to the 9 deg$^2$ area of \citeauthor{IH95}.  Were the
\citeauthor{paperII} Carina density profile (Fig.\ 1) taken at face
value, it is conceivable that as much as half of the \citeauthor{W03}
``background density'' --- estimated from the edge of their survey field
--- is from Carina itself.  \citeauthor{W03} do mention the possibility
that if extended dSph populations are large enough to extend to the
limit of their survey area then they would overestimate the level of
their background.  As we show in \S3, Carina stars do exist to at least
$40\arcmin$ along the major axis; this circumstance alone has probably
significantly contributed to an overestimate of the \citeauthor{W03}
background density, and could explain why their deeper survey did not
see an extratidal feature reported by the shallower, but larger area
\citeauthor{IH95} survey.

Yet another difference between the surveys relates to the relative
magnitude errors.  \citeauthor{IH95} estimate the possibility of 0.1 mag
large-scale systematic errors in their photographic data; they do not
give information on the random errors but at least these are expected to
be relatively uniform over their photographic plates.  In contrast,
mosaiced surveys of CCD data are notoriously inhomogeneous due to seeing
and transparency variations during observation.  It is perhaps significant that
\citeauthor{W03} report 0.1 mag uncertainties in absolute brightness for
their {\it brightest} stars ($V<20$) but up to 0.3 mag uncertainties
among their fainter stars.  Although the photometry of their CCD frames
has been tied together with stars in overlap regions, this does not
account for differences in the level of \citet{edd13} bias from CCD
frame to CCD frame, which, with 0.3 mag errors and a steeply rising
background source count, could contribute significant excess large-scale
``noise'' on top of the Poissonian fluctuations.  Depending on the
overall uniformity of their CCD frames, the effect of variable Eddington
bias may be inconsistent with the conclusion by \citeauthor{W03} that
``the precision of the photometry is not critical for this work.''

In the end it is not absolutely clear why \citeauthor{IH95} detected an
``extratidal excess'' in Carina but \citeauthor{W03} could not confirm a
break to shallower slope in the Carina profile when they probed down to
Carina main sequence magnitudes.  It should be noted that, in spite of
their repeated finding of extratidal detections around Galactic dSphs,
\citeauthor{IH95} do discuss background underestimation as one possible
explanation for these (perhaps false) detections in their data, although
they argue that this is unlikely for most of their dSphs.  The intent of
the above detailed comparison of \citeauthor{IH95} to \citeauthor{W03}
is to demonstrate some of the pitfalls of the difficult and tedious work
of deriving density profiles from starcounts and, moreover, to question
the assumption in this particular kind of survey work that ``deeper is
necessarily better''.

\subsubsection{``Filtered'' Starcount Studies of the Carina dSph}

\citeauthor{W03} have used deeper imaging as one method to increase the
dSph signal with respect to the background noise under the operative
philosophy that the dSph signal rises faster than the increase in
contributed background noise as one probes to fainter magnitudes.  An
alternative approach to increasing the $S/N$ of diffuse dSph features is
to work hard on beating down the size of the contributed background
noise by identifying and selectively weeding out those sources most
likely to be unrelated to the dSph.  If multifilter data are in hand,
one can use the fact that dSphs have well-defined loci in the
color-magnitude diagram to eliminate a large fraction of the background
sources inconsistent with membership in the dSph.  The method of
counting stars lying in well-chosen regions of the color-magnitude
diagram (CMD) where the target signal-to-background noise ratio is
optimized has been applied to the search for tidal tails around globular
clusters by \citet*{gri95,gri96,LMC00,roc02}, and \citet{ode01a,ode03},
around dSph galaxies by \citet{kle98} and \citet{pia01}, as well as to
searches for tidal tails in the Galactic halo
\citep{md01,md02,bel03,iba03}.  In a variant of this technique,
\citet*{KSH96} fitted the CMDs of extratidal fields around the Carina
dSph with combinations of ``Carina'' and ``background'' CMD templates,
and found a radial profile break population with a density roughly
consistent with those found by \citeauthor{IH95} and
\citeauthor{paperII}.  Because this information is not available from
their paper, we cannot include the \citeauthor{KSH96} relative
Carina-to-background density in Figure 1, but it is expected to be
better than that of \citeauthor{IH95}.  More recently, \citet{mon03}
have reported a ``shoulder'' in the density distribution of stars
selected from regions in the Carina CMD meant to be dominated by old
Carina stars (the RR Lyrae, BHB and subgiants), and suggest it may be
related to the predicted \citet{JSH99} radial profile ``break'' in
tidally disrupting systems; however, this \citeauthor{mon03} break
occurs well inside the King radius (4-6 arcmin from the center of
Carina) and its relation to the data from other surveys shown in Figure
1 is unclear.  On the other hand, in a more recent contribution,
\citet{mon04} report the likely presence of the old Carina MSTO in a CMD
for a field located just outside of the Carina tidal radius, a result
that supports the notion of a true break population there.

To improve the signal-to-noise ($S/N$) of the tenuous extratidal
features even further, \citeauthor{paperII} added to the optimized CMD
filtering technique an additional strategy to identify and remove
residual ``noise'' that happens to fall within the selected regions of
the CMD.  Because this residual ``background noise'' is actually
dominated by foreground Galactic disk dwarf stars, \citeauthor{paperII}
relied on a photometric method that can discriminate them from Carina
giant stars: The Washington $M,T_2+DDO51$ filter technique
\citep{gei84,paperI} relies on the surface gravity sensitivity of the
Mgb+MgH spectral features near 5150\AA\ in late G- and K-type stars (the
feature is secondarily sensitive to metallicity).  The result of this
two step filtering technique is a drastic reduction of the background
level and the resultant detection of a very obvious break population in
the distribution of Carina giant stars (Fig.\ 1) which is consistent
with that reported by both \citeauthor{IH95} and \citet{KSH96}.  A
benefit of the \citeauthor{paperII} analysis is that it goes beyond a
mere {\it statistical} measurement of the Carina radial profile; rather,
it endeavors to identify {\it precisely which} stars are members of the
dSph, and these can be spectroscopically targeted for a straightforward
check of the veracity of the derived photometric profile (see \S3).  The
technique also supplies a relatively pure target list of the most
accessible stars to use for study of the dSph's dynamics (Mu\~noz et
al., in preparation).  Figure 1 shows that of all the photometric Carina
studies to date, the \citeauthor{paperII} survey is the {\it only one}
for which the identified break population is at a density {\it greater}
than its subtracted background and {\it by almost an order of
  magnitude}, in contradistinction to the other surveys for which the
inverse (or worse) is true.

Said another way, the \citeauthor{paperII} background would have to be
underestimated by {\it almost a factor of ten} in order to erase the
detected break population in \citeauthor{paperII}.  Moreover, the error
in that estimated background would have to show a {\it systematic radial
  trend away from the center of Carina} in order to create the false
signal of a power-law decline in the dSph outside the nominal King
limiting radius.  While it would seem difficult for any analysis to have
made an error in measurement both so incredibly large as well as so
remarkably unfortunate as to have an {\it inverse} power-law radial
variation, this is exactly the position taken by
\citet*[][``MOMNHDF'']{MOMNHDF}, who, through a reanalysis of the
\citeauthor{paperII} data, conclude that the detected break population
is entirely an artifact of photometric errors.  However, as we now show,
the \citeauthor{MOMNHDF} analysis contains several errors and mistaken
assumptions about the \citeauthor{paperII} study that leads them to this
incorrect conclusion.

\section{Assessments of the \citeauthor{paperII} Contamination Level}

\subsection{Review of Goals and Conclusions of \citeauthor{paperII}}

\citeauthor{paperII} used the $M,T_2,DDO51$ technique to identify an
extended distribution of giant star candidates beyond the nominal
``tidal radius'' of the Carina dSph galaxy, a stellar excess also
previously reported by \citeauthor{IH95} and \citet{KSH96}.  As
discussed above and in these cited references, the existence of such
break populations may have profound implications for the outer
structure, dark matter content, disruption history, and/or perceived
star formation histories of satellite dSph galaxies, as well as the
structure and formation of the Milky Way halo.  \citeauthor{paperII}
discussed {\it several} physical explanations for the apparent extended
distribution of stars found beyond the nominal Carina tidal radius,
including the {\it possibility} that Carina is losing about 27\% of its
mass per Gyr.  As stated in Paper II, each of the various possible
explanations ``has a distinct, kinematical signature that would be
recognizable with an appropriate radial velocity survey.''  Without
radial velocity data, not only were such speculations about the {\it
  meaning} of the Carina radial structure beyond verification, but the
actual analysis of the Carina structure itself was necessarily based, in
large part, on ``extratidal'' star {\it candidates} around which it was
felt that a strong case for reliability had been built --- but {\it
  candidates} nonetheless.  Because of an inability to have made
significant progress with spectroscopic testing of these candidates,
despite repeated attempts over five years, the \citeauthor{paperII}
analysis of Carina relied in part on statistical arguments, including,
for example, in the assessments of sources of potential contamination of
the giant candidate sample.  Thus, \citeauthor{paperII} made a first
attempt, via comparison of background stellar densities in regions of
color-magnitude space adjacent to that region used to select Carina
giant star candidates, to account for two sources of potential
contamination: (1) field giants and metal-poor subdwarfs having similar
$M-DDO51$ colors to the selected Carina giant sample, and (2) non-Carina
stars errantly scattered into the photometrically-selected sample due to
random errors in the photometry.  However, \citeauthor{paperII} did
present a small-scale spectroscopic test of the accuracy of the
photometric dwarf/giant separation: For a proxy sample of 27 Carina
region stars for which spectroscopic data existed a 100\% accuracy in
the dwarf/giant classifications was found, and among these 27 stars were
three newly discovered giants outside the nominal Carina tidal radius.

\subsection{Review of Goals and Conclusions of \citeauthor{MOMNHDF}}

Subsequently, \citeauthor{MOMNHDF} re-addressed the issues of subdwarf
contamination and photometric errors in Washington$+DDO51$ surveys;
their analysis relied on Monte Carlo simulation of photometric error
spreading in the Washington$+DDO51$ filter color-color diagram for the
latter effect.  \citeauthor{MOMNHDF} include a reexamination of the
contamination rate in the Paper II study of Carina, and reach
dramatically different conclusions about it: As summarized in their
Figure 13, \citeauthor{MOMNHDF}'s results suggest that the {\it entire}
sample of ``extratidal'' Carina giant candidates identified in
\citeauthor{paperII} can be accounted for as contamination by
misidentified subdwarfs and by photometric errors that scatter Galactic
disk dwarf stars from the ``dwarf'' region of the $(M-T_2, M-DDO51)_o$
diagram into the ``giant'' region.  In addition, \citeauthor{MOMNHDF}
discount the \citeauthor{paperII} spectroscopic test as being relevant
to assessing the veracity of that work, since they consider at least two
of the spectroscopically confirmed ``extratidal'' giants as
insufficiently outside the tidal radius to be significant (when they
account for errors in the establishment of that radius by
\citeauthor{IH95}\footnote{The concordance of the various radial
  profiles and King function fits shown in Figure 1 suggests that the
  true uncertainty in the overall fitted profiles may actually be
  reasonably small.}).

\citeauthor{MOMNHDF} conclude: {\it{``If one wishes to make statistical
    corrections for the number of bogus giants caused by photometric
    errors, it is important to quantify accurately the photometric
    errors in the data, both in terms of the average error and the shape
    of the distribution.''}}  This warning, as well as the general
themes of the \citeauthor{MOMNHDF} paper, emphasize, appropriately, the
caution one must have in \textit{any} study making use of
photometrically selected samples.  Ironically, however,
\citeauthor{MOMNHDF} themselves have inaccurately quantified both the
``average error and the shape of the [Paper II error] distribution'', so
that they have greatly overestimated the number of potential
contaminants in the \citeauthor{paperII}, candidate Carina giant sample.
Additional simplifications and incorrect assumptions in the
\citeauthor{MOMNHDF} analysis have further inflated their estimated
Paper II contamination levels, as demonstrated in the following
subsections.  We first reexamine the \citeauthor{MOMNHDF} numerical
technique (\S2.3) and then introduce an alternative, and we believe more
accurate, \textit{analytical} method for \textit{a posteriori}
assessment of contamination levels (\S2.4).  However, {\it through both
  methods} we find not only much more modest levels of dwarf
contamination (at levels near those calculated in the original
\citeauthor{paperII} analysis via its {\it third}, independent
assessment method), but that in seeking to be conservative,
\textit{\citeauthor{paperII} might actually \emph{underestimate} the
  density of extratidal Carina giant stars}.

\subsection{Numerical Simulation of Contamination Level}

An \textit{ideal} numerical analysis of the \citeauthor{paperII}
photometric survey along the lines of that approximated by
\citeauthor{MOMNHDF} would proceed by (1) invoking a ``truth''
distribution, $\mbox{\boldmath{$X$}}$, of points
$\mbox{\boldmath{$x$}}_i$ representing both Milky Way and Carina stars
in the {\it three}-dimensional space $(M-T_2,M-DDO51,M)_o$ and (2)
applying random deviates, $\mbox{\boldmath{$\xi$}}_i$ (appropriately
matched to the measured error distribution in each dimension) to create
a new distribution, $\mbox{\boldmath{$Y$}}$, of perturbed points
$\mbox{\boldmath{$y$}}_i = \mbox{\boldmath{$x$}}_i +
\mbox{\boldmath{$\xi$}}_i$, in the three-dimensional space.  After
perturbation of the truth sample by one level of error, (3) the number
of stars that cross the three-dimensional ``Carina-giant candidate''
selection boundaries of \citeauthor{paperII} (in {\it either} direction)
may be evaluated. With Monte Carlo methods for selecting the random
deviates, (4) the process may be repeated multiple times to evaluate an
expected mean level of dwarf contamination of the ``Carina-giant
candidate'' sample.

Unfortunately, a number of unavoidable factors make it difficult to
realize this ``ideal'' methodology, but \citeauthor{MOMNHDF} have in
addition made some (avoidable) simplifications that have a critical
impact on their assessment of Paper II.  Here we discuss various points
pertinent to a \citeauthor{MOMNHDF}-like numerical simulation of the
Paper II photometric errors:

$\bullet$ \textit{Picking an appropriate truth sample.}  To implement
the ``ideal'' numerical simulation of the effects of errors on the
experiment requires the adoption of a suitable representation of an
errorless data distribution.  Unfortunately, the truth distribution
$\mbox{\boldmath{$X$}}$ is ultimately unknown, and may only be
approximated by an observed distribution,
$\mbox{\boldmath{$X^\prime$}}$, where each point is already perturbed by
one level of error, $\mbox{\boldmath{$x$}}_i \rightarrow
\mbox{\boldmath{$x$}}_i + \mbox{\boldmath{$\xi$}}_i^\prime$.  Thus, any
numerical experiment where the number of error-perturbed dwarf stars
becoming giant contaminants is evaluated that starts with an
\textit{observed} dataset must be acknowledged \textit{a priori} to
overestimate the amount of that contamination.  Clearly, the smaller the
$\mbox{\boldmath{$\xi$}}_i^\prime$, the more reliable the approximation
to the truth distribution.  For the simulation they show in their Figure
7, \citeauthor{MOMNHDF} have taken a ``dwarf locus'' from their own
survey, after imposing a severe 0.02 magnitude error limit in each
magnitude, so that, in general, $\overline{|\mbox{\boldmath{$\xi$}}_i'|}
<< \overline{|\mbox{\boldmath{$\xi$}}_i|}$ for each star in their
simulation.  This seems a reasonable simplification for their simulation
only because \citeauthor{MOMNHDF} have assumed
$\overline{|\mbox{\boldmath{$\xi$}}_i|} = 0.10$ mag; however, as we show
below, this is not at all typical of the errors in the
\citeauthor{paperII} sample, which are more like 0.033 mag.  For a large
fraction of the stars in the \citeauthor{paperII} sample, the
$\overline{|\mbox{\boldmath{$\xi$}}_i'|}$ \citeauthor{MOMNHDF} adopted
in their Figure 7 simulation are relatively close to the
$\overline{|\mbox{\boldmath{$\xi$}}_i|}$ of \citeauthor{paperII}.  The
situation is, however, even worse for the \citeauthor{MOMNHDF} analysis
of ``giants kept/dwarfs gained'' (their Figs.  8 and 9) since here the
authors have imposed {\it no} error limit on their ``zero error'' input
data set.

Even were one to adopt an essentially ``zero-error'' (but observed)
distribution of dwarfs (or any stars\footnote{Note that
  \citeauthor{MOMNHDF}'s analysis excludes two parts of the ``truth
  distribution'': field giants and actual Carina stars.  The latter need
  to be addressed, as we do below, in the context of scatter {\it out}
  of the selection process.  A small number of field giants might also
  be expected to contribute a share of potential contaminants.  Both our
  analysis in \citeauthor{paperII} as well as that presented here in \S3
  account for these extra contaminants not modeled by
  \citeauthor{MOMNHDF}.})  to represent the truth distribution in one
direction of the sky, a second problem arises in that the truth
distribution is not a constant around the sky.  For example, the shape
of the ``dwarf locus'' is a function of Galactic position because of its
dependence on the metallicity distribution of those stars \citep[see,
for example, Figure 2 of][]{paperI}.  \citeauthor{MOMNHDF} have taken
their ``zero-error'' distribution from their own survey fields with
latitude ranging from $25\arcdeg \le b \le 72\arcdeg$; however, since
Carina is at $b=-22\arcdeg$, one would expect a proper ``truth''
distribution of dwarf stars for the Carina field to contain a
metallicity distribution skewed towards higher abundance disk stars than
does a high latitude sample.  Because higher metallicity dwarfs lie
further from the \citeauthor{paperII} giant/dwarf separation in the
two-color diagram (2CD) than do lower metallicity dwarfs, it is likely
that \citeauthor{MOMNHDF}'s ``zero-error'' dwarf locus will admit more
contaminants into the giant region than would a dwarf locus appropriate
to the Carina line-of-sight.

$\bullet$ \textit{Using a three-dimensional analysis}.  The
\citeauthor{MOMNHDF} discussion focuses on their simulation of error
propagation in the $(M-T_2,M-DDO51)_o$ plane.  However,
\citeauthor{paperII} selected a star to be a ``Carina RGB candidate''
based {\it not only} on it having a position in the $(M-T_2, M-DDO51)_o$
plane expected for giant stars, but {\it also} a position in the
$(M-T_2, M)_o$ color-magnitude plane commensurate with probable
association with the Carina dSph.  While the first criterion used was
generally more liberal than that used by the ``spaghetti group'' in
their own selection of giant candidates (see \citeauthor{MOMNHDF}), the
second criterion was designed to be so conservative in disallowing
potential Carina-RGB candidate contaminants that it more than makes up
for the relatively more vulnerable color-color criterion, and it is a
critical aspect of our process.  For example, it is not clear why the
\citeauthor{MOMNHDF} simulation of \citeauthor{paperII} permits stars as
blue as $(M-T_2)=0.5$ and why the authors focus on this ``blue extension
[that] causes problems'' since application of the various Paper II CMD
selection criteria\footnote{The most generous of the
  \citeauthor{paperII} CMD selection criteria, for a magnitude limit
  $M=20.8$, are illustrated in Figure 5a below.  Three brighter samples
  were also analyzed in Paper II, where the lower limit of the CMD
  selection criterion was raised to $M=20.3$, 19.8 and 19.3, a
  progression that increasingly restricts the allowed $(M-T_2)$ color
  range.}  significantly reduces or entirely eliminates this ``blue
extension''.  Indeed, for the \citeauthor{paperII} $M<19.3$ sample {\it
  the ultimate $(M-T_2)$ color selection is at least as strict, if not
  more so, than that applied by \citeauthor{MOMNHDF} in their own
  survey}.  Failure to incorporate the true \citeauthor{paperII} color
limits has substantially inflated the number of ``MOKP [\citet{paperI}]
bogus giants'' demonstrated in \citeauthor{MOMNHDF}'s Figure 7, for
example.  It is the {\it three-dimensional} aspect of the
\citeauthor{paperII} selection criterion, i.e., that a star must land in
a very restricted volume of $(M-T_2, M-DDO51, M)_o$ space, that makes
its ultimate selection of candidates so conservative.  Moreover, as
shown below, the larger mean density of stars inside this parameter
space selection volume compared to the density just outside of it means
that it is statistically more likely for bona-fide Carina RGB stars to
be scattered {\it out of} the small selection volume in parameter space
than for non-Carina-RGB stars to enter it, even when photometric errors
as large as 0.1 mag per filter (those invoked and studied by
\citeauthor{MOMNHDF}) are allowed.

$\bullet$ \textit{Assuming Gaussian deviates.}  As pointed out by
\citeauthor{MOMNHDF}, the detailed {\it shape} of the error distribution
can play an important role in the outcome of a numerical simulation.  In
the face of no information to the contrary, it is common (and generally
easy) to assume that the distribution of errors is Gaussian.  Obviously,
the wings of a platykurtic error distribution will produce more spurious
contaminants, while a leptokurtic distribution will yield fewer
problems.  We have checked the shape of the \citeauthor{paperII} error
distribution through an analysis of the errors in the magnitude
distributions of artificial stars and find that the \citeauthor{MOMNHDF}
assumption of a Gaussian shape to the Paper II error distribution is
valid.

$\bullet$ \textit{Assuming a proper-sized distribution for the Gaussian
  errors.}  A fundamental problem with \citeauthor{MOMNHDF}'s critique
of \citeauthor{paperII} is that the level of error they have adopted to
represent the photometric sample in the evaluation of the number of
potential contaminants is too large.  \citeauthor{MOMNHDF}'s criticism
of the Carina giant candidate selection might be well-founded if indeed
the \textit{typical} photometric errors were as large as 0.1 mag;
however, 0.1 mag is the \textit{magnitude cut-off} applied in the
\citeauthor{paperII} analysis, and can hardly be considered the
\textit{typical} magnitude error.  Obviously, attributing the
\textit{worst} photometric error (0.1 mag per filter, or 0.14 error per
color) to the \textit{entire} sample of \citeauthor{paperII} stars
grossly exaggerates the degree of smearing of the dwarf locus (as
represented in \citeauthor{MOMNHDF}'s Figs.\ 4 and 13) and results in a
greatly overestimated number of contaminants scattered into the giant
selection region in $(M-T_2, M-DDO51)_o$ space.  \citeauthor{MOMNHDF}'s
conclusions that the \citeauthor{paperII} ``extratidal'' population may
be \textit{entirely} due to dwarf star contamination, which they obtain
from the simulation results summarized in their Figure 13, can only be
reached after applying this \textit{maximum} photometric error in the
\citeauthor{paperII} sample to \textit{all} stars.

{\it Even then} the {\it apparent} convergence of their worst case, 0.15
mag color error models to a 100\% contamination within the extratidal
Carina giant sample shown in \citeauthor{MOMNHDF}'s Figure 13 is
incorrect because \citeauthor{MOMNHDF} have undercalculated/mislabeled
the density of extratidal giants that \citeauthor{paperII} identified as
100 deg$^{-2}$, when the actual density that \citeauthor{paperII}
reported is 124 deg$^{-2}$.  The mistakenly reported lower density value
likely derived from having used the \textit{entire} area of the
\citeauthor{paperII} survey to calculate the \textit{extratidal} density
(i.e., 1.16 deg$^2$), and neglecting to exclude the area in the
intratidal region (0.22 deg$^2$).  This 23\% underestimate also affects
the background (i.e., contaminant) density that they report from
\citeauthor{paperII}: The actual ``MOPKJG [\citeauthor{paperII}]
background estimate'' is 27.9 deg$^{-2}$, not the 22.5 deg$^{-2}$
\citeauthor{MOMNHDF} show.

As shown by Figure 2, the reality of the \citeauthor{paperII} error
distribution is far less pessimistic than the strawman, ``photographic
quality'', error distribution that \citeauthor{MOMNHDF} have criticized:
The median errors in the deepest, $M=20.8$ sample of
\citeauthor{paperII} are at $\sigma_M = 0.033$ mag,
$\sigma_{M-T_2}=0.045$ and $\sigma_{M-DDO51}=0.050$\footnote{That the
  error distributions of all four magnitude-limited samples shown in
  Figure 2 are similar is due to the fact that in \citeauthor{paperII},
  survey subregions having incompleteness at the imposed magnitude limit
  were dropped from the analysis, a step in the Paper II process that
  eliminates at each magnitude limit those CCD data contributing the
  most offensive errors.  Note also that all of our quoted random errors
  include some additional inflation (typically expected to be
  $\lesssim10\%$) due to the the propagated contribution of possible
  {\it{systematic}} shifts due to errors in the coefficients in the
  photometric transformation equations.  Such systematic shifts play
  less a role in inducing contamination since they more or less affect
  all stars similarly.}.  Therefore, it is worthwhile assessing what
\citeauthor{MOMNHDF} would have concluded from their own analysis had
they accounted for the actual error distribution in the
\citeauthor{paperII} data: For the {\it actual} median Paper II color
error of $\sim 0.0475$ mag, the \citeauthor{MOMNHDF} model results as
presented in their Figure 13 yields an expected number of ``interloper
dwarfs'' among the extratidal Carina giant candidates much smaller than
\citeauthor{MOMNHDF} have implied, and, moreover, very near what the
original \citeauthor{paperII} analysis estimated them to be --- i.e.
about 30 deg$^{-2}$!

However, the latter simple comparison ignores the potentially
deleterious effects of the small number of stars with larger than
typical photometric errors in the asymmetric wing of the error
distribution --- the problem objects more similar to the kind (but not
proportion) \citeauthor{MOMNHDF} simulated.  Thus a better assessment of
what \citeauthor{MOMNHDF}'s model would predict in the case of the true
photometric error distributions comes by finding the expectation value
of the product of their Figure 13 interloper function with our Figure 2
error distributions.\footnote{Without their ``truth'' distribution, we
  cannot rerun the \citeauthor{MOMNHDF} model from scratch.}  Because
\citeauthor{MOMNHDF} present results (their Fig.\ 13) only in the case
for equal errors in all magnitudes, whereas the actual data have varying
errors in each dimension of $(M-T_2, M-DDO51, M)_o$ space, for our
calculation we adopt from the \citeauthor{MOMNHDF} Figure 13 a
fractional interloper expectation for each star inferred from the
abscissa point given by the geometric mean of the axial radii of the
error ellipsoid, $ 2^{1/2}(\sigma_M \sigma_{T_2} \sigma_{DDO51})^{1/3}$,
where the $2^{1/2}$ comes from the fact that \citeauthor{MOMNHDF}'s
Figure 13 presents results in terms of color, not magnitude, error.  For
the $M=20.8$ sample (the case presented in \citeauthor{MOMNHDF}'s Figure
13), we find that \citeauthor{MOMNHDF}'s model with the actual error
distribution of the \citeauthor{paperII} sample gives a density of
photometric error contaminants of 22 deg$^{-2}$.  When one adds in the
14 predicted metal-poor subdwarfs from their model, one gets 36 total
contaminants -- again, rather similar to the total contamination level
of 28 deg$^{-2}$ calculated in \citeauthor{paperII}.  The relatively
close agreement between the \citeauthor{MOMNHDF} model results using the
\textit{proper} \citeauthor{paperII} error distribution and what we
derived for the net contamination of the Carina giant candidate sample
from the very different analysis in \citeauthor{paperII} suggests that
the adoption of inflated photometric uncertainties is the predominant
shortcoming in \citeauthor{MOMNHDF}'s analysis.

$\bullet$ \textit{Measuring the missed detection rate.}
\citeauthor{MOMNHDF} concentrated on the number of ``bogus'' sources
entering into the Carina giant sample.  However, without also including
a distribution of Carina giants in their simulation,
\citeauthor{MOMNHDF} did not address the possibility of sample flux in
the other direction --- i.e., a loss of true Carina stars from the
sample.  For stars near the three-dimensional selection boundary, the
volume of parameter space available to stars {\it leaving} the selection
volume is larger than for stars that would scatter {\it into} the
comparatively small selection volume.  Thus, \citeauthor{MOMNHDF}'s, as
well as {\it our} previous, estimates of the degree to which the Paper
II extratidal densities are inflated by photometric errors ignores a
potentially significant countereffect that {\it reduces} measured Carina
densities relative to the background (indeed, even increasing the
background level as estimated by the methodology of
\citeauthor{paperII}), making the the derived extratidal giant densities
more conservative.

\subsection{Analytical Estimation of Contamination Level}

\citeauthor{MOMNHDF}'s evaluation of sample contamination employs a {\it
  model} of the data to numerically simulate the effects of photometric
errors.  In this subsection we describe an alternative, analytical,
\textit{a posteriori} approach to ascertaining sample contamination
levels by using \textit{the data themselves}.  With it we reaffirm the
independent \citeauthor{paperII} analysis of the level of contamination
in our final ``Carina giant candidate'' sample, which is significantly
less than proposed by \citeauthor{MOMNHDF}.

The number of stars scattered into our three-dimensional $(M-T_2,
M-DDO51, M)_o$\footnote{For the remainder of this section, we drop the
  subscript ${}_o$ for clarity (though in fact we work with the
  dereddened photometry throughout).} Carina giant selection due to
photometric errors can be assessed by considering the probability
distribution function for the colors and magnitudes of individual stars.
What we calculate for each giant candidate is the \textit{posterior}
probability that it belongs inside the giant selection region.  This
calculation assumes that the giant selection region that we have
employed separates giants from dwarfs with perfect accuracy, which is
the simplest assumption one can make for now; when additional
information becomes available (e.g., through an empirical determination
of the true contamination rate via spectroscopic observations of a large
sample of our giant candidates) the assumption about the accuracy of our
giant selection region can be adjusted (see \S2.4.1).

Each star identified as a candidate Carina giant by our combined
color-color and color-magnitude selection process has an associated
photometric error in each filter, $\sigma_{M}$, $\sigma_{T_2}$, and
$\sigma_{DDO51}$.  These quantities can be used to compute a covariance
matrix for the magnitude, $M$, and the colors $M-T_2$ and $M-DDO51$,
which represents the total error in the position of a star within the
three-dimensional color-color-magnitude space.  The covariance matrix
can be written as
\begin{equation}
[C] = \left[
\begin{array}{ccc}
\sigma^2_M & \sigma^2_M & \sigma^2_M \\
\sigma^2_M & \sigma^2_M+\sigma^2_{T_2} & \sigma^2_M \\
\sigma^2_M & \sigma^2_M & \sigma^2_M + \sigma^2_{DDO51}
\end{array}
\right],
\end{equation}
where only the terms involving $\sigma^2_M$ survive in the off-diagonal
terms because the errors in different magnitude terms are uncorrelated.
 
Assuming that the photometric errors in each filter have a Gaussian
probability distribution, then the likelihood that each star, $i$, has
an $(M-T_2)^{\prime}$ color, $(M-DDO51)^{\prime}$ color, and
$M^{\prime}$ magnitude different from its measured values,
$(M-T_2)_{i}$, $(M-DDO51)_{i}$, and $M_{i}$, is
\begin{equation}
L(M-T_2^{\prime},M-DDO51^{\prime},M^{\prime}) = 
\frac{1}{(2\pi)^{3/2} \, \sigma_{M} \, \sigma_{T_2} \, \sigma_{DDO51}}
\exp\left(-\frac{1}{2}x^2\right),
\label{gprob}
\end{equation}
where
\begin{equation}
x^2 \equiv [\Delta]^T[C]^{-1}[\Delta],
\end{equation}
and $[\Delta] \equiv [M' - M_i, (M-T_2)' - (M-T_2)_i, (M-DDO51)' -
(M-DDO51)_i]$.  Using Bayes' theorem, the probability that each star is
either a giant or a contaminant becomes
\begin{equation}
P([M-T_2]^{\prime},[M-DDO51]^{\prime},M^{\prime}) = \frac{L([M-T_2]^{\prime},[M-DDO51]^{\prime},M^{\prime}) 
* P_0([M-T_2]^{\prime},[M-DDO51]^{\prime},M^{\prime})}{D}
\label{prob},
\end{equation}
where $P_0([M-T_2]^{\prime},[M-DDO51]^{\prime},M^{\prime})$, known as
the {\em prior}, is the {\em a priori} probability (based on the
information $I$) that we would consider each star a contaminant, and $D$
is a normalization factor chosen such that the integral of
$P([M-T_2]^{\prime},[M-DDO51]^{\prime},M^{\prime})$ over all possible
values of $M$, $M-DDO51$ and $M-T_2$ is one.

\subsubsection{An Aside About Prior Probabilities}

In order to evaluate equation~(\ref{prob}) above, we must write an
expression for the prior,
$P_0([M-T_2]^{\prime},[M-DDO51]^{\prime},M^{\prime})$.  The proper
choice of this quantity is often a subject of heated debate in
discussions of Bayesian methods.  In fact, a properly chosen prior
should rarely make a significant difference in the final results.  To
see why this is so, consider the strategies by which a prior might be
chosen.

First, we might choose an ``uninformative prior''; that is, a prior that
makes weak assertions about the \emph{a priori} distribution of the
variables being measured.  Priors typically used for this purpose
include the uniform prior, $P(X) = \mbox{\rm constant}$ and the
Jeffreys' prior\footnote{Both of these priors are unnormalized.  In
  practice this is not a problem, as the product of likelihood and prior
  has a finite integral.}, $P(X) \propto 1/X$.  When using a prior of
this sort, the posterior probability will be dominated by the
contribution of the likelihood function.  In essence, we ``forget'' our
a priori assumptions once we have actual measurements in hand.

Alternatively, we might choose an ``informative prior.''  This would be
especially appropriate, for example, if there were an accepted value
derived from previous experiments for the quantity under investigation.
In this case the prior and the likelihood \emph{ought} to agree.  If
they do, then the effect of the prior is to tighten the confidence
intervals around the peak of the posterior distribution; this is
equivalent to a meta-analysis of the data from all of the experiments.
If the prior and the likelihood do not agree, then we should probably
not be applying equation~(\ref{prob}) blindly; we should instead figure
out the reason for the discrepancy.

For this study the probability density in the color-color-magnitude
space is probably not uniform, but we are reluctant to choose an
informative prior unless it can be rigorously defended, which in
practice means we limit our use of informative priors to meta-analysis.
Accordingly, we adopt the uniform prior for these calculations, and
equation~(\ref{prob}) becomes
\begin{equation}
P([M-T_2]^{\prime},[M-DDO51]^{\prime},M^{\prime}) \propto
L([M-T_2]^{\prime},[M-DDO51]^{\prime},M^{\prime}),
\end{equation}
with the constant of proportionality set by the normalization condition.

\subsubsection{Contamination Evaluation}

Let $R$ denote the giant selection region of ($M-T_2$, $M-DDO51$, $M$)
space, and $\tilde{R}$ its complement.  Then the probability, $P_i$,
that a star, $i$, belongs in $\tilde{R}$ and not in $R$ is
\begin{equation}
P_{i} = \int_{\tilde{R}} P([M-T_2]^{\prime},[M-DDO51]^{\prime},M^{\prime}) 
\: d(M-T)' \: d(M-D)' \:  dM'.
\label{pint}
\end{equation}
The integral in equation~(\ref{pint}) is difficult to evaluate
analytically because the boundaries are irregularly shaped;
consequently, we use a Monte Carlo integration technique \citep[e.g.,][
\S7.6]{numrec}, in which $\int f dV \approx V \langle f \rangle$, where
$f$ is a function and $V$ is the parameter space volume over which we
are integrating.  We generate integration sample points using a
quasi-random sequence (\citeauthor{numrec}, \S7.7).  Although the
integral extends formally to infinity in the ($M$, $M-T_2$, $M-DDO51$)
space, for practical reasons we impose a bounding box, which we make
large enough that $P_i$ is negligible along the boundary for all stars
in the sample.  The integration samples span the entire ($M-T_2$,
$M-DDO51$, $M$) space within the box, but only those in $\tilde{R}$ are
allowed to contribute to the integral.  Then, the volume of $\tilde{R}$,
$V_{\tilde{R}}$, is given by $(N_{\tilde{R}}/N_{\rm tot})V_{\rm tot}$, where
$(N_{\tilde{R}}$ is the number of integration samples within
$\tilde{R}$, and $N_{\rm tot}$ is the total number of samples.  The
expectation value for the number of $\tilde{R}$ stars that appear in $R$
due to photometric error is then
\begin{equation}
\langle N_{c} \rangle = \sum_{i=1}^{n} P_{i}.
\label{ncontam}
\end{equation}

There are two sources of uncertainty in $\left\langle N_c\right\rangle$.
The first is the error in the approximation to the integral in
equation~(\ref{pint}).  The Monte Carlo technique used to evaluate the
integral is a numerical approximation, and the sampling error,
$\sigma_{MC,i}$, in this approximation for each star $i$ is given by
$\sigma^2_{MC,i} = V^2 (\langle f_i^2 \rangle - \langle f_i \rangle^2) /
N $.  The total variance in $\left\langle N_{c} \right\rangle$ from
these errors is then $\sigma^2_{MC} = \sum_i \sigma^2_{MC,i}$.

The second source of uncertainty is the statistical uncertainty inherent
in any random experiment; viz., the number of contaminants actually
observed will not be exactly $\left\langle N_{c} \right\rangle$, but
will instead have some probability distribution peaked at $\left\langle
  N_{c} \right\rangle$.  We can calculate the variance of this
distribution by considering the two possibilities for each star found in
the selection region.  Either the star is a true Carina giant ($c=0$),
or the star is a contaminant ($c=1$).  If the star's contribution to
$\left\langle N_{c} \right\rangle$ is $\mu = P_i$, then the variance
contributed is
\begin{equation}
\sigma^2_i = \sum_{c=0,1} \left[(c - \mu)^2 P_i(c)\right],
\end{equation}
or
\begin{equation}
\sigma^2_{stat} = \sum_i \sigma^2_i = \sum_i P_i (1-P_i),
\end{equation}
since $P_i(c=0) = 1-P_i$ and $P(c=1) = P_i$.
Finally, the combined uncertainty from both
effects is $\sigma_{\langle N_c
\rangle} = \sqrt{\sigma_{MC}^2 + \sigma_{stat}^2}$.

The statistical uncertainty is fixed for any particular sample; however,
the error in the numerical approximation is a function of how many
points are sampled in the integration, and can be reduced to
insignificance.  The Monte Carlo integration was carried out with
10,000,000 integration samples, so that $\sigma_{\rm MC} \lesssim 1/4
\sigma_{\rm stat}$ in the majority of cases.  Test calculations were run
with the number of integration samples ranging from 25,000--10,000,000
and with integration limits extending to $12 < M^{\prime} < 25$, $-2 <
(M-T_2)^{\prime} < 6$, and $-1.0 < (M-DDO51)^{\prime} < 1.0$.  Other
than the expected decrease in $\sigma_{\rm int}$ due to the increase in
integration samples, the resulting number of contaminants remained
constant through the various trials; thus, the contamination estimates
are insensitive both to the number of points used in the integral
approximation and to the limits of integration.

In Table~\ref{contamprob} we report the expected number of contaminants
found in various subsamples of candidate giant stars drawn from the
catalogue of all stars observed in our survey of the Carina dSph.
Various subsamples are used to determine whether the rate of
contamination is significantly larger at faint magnitudes or in the
extratidal region than it is at bright magnitudes or in the core of the
galaxy.  The integration was carried out over a finite region of
parameter space that included all values of $(M-T_2)^{\prime}$,
$(M-DDO51)^{\prime}$, and $M^{\prime}$ that might possibly contribute to
the integral: The adopted ranges are $14 < M^{\prime} < 23$, $-1 <
(M-T_2)^{\prime} < 5$, and $-1.0 < (M-DDO51)^{\prime} < 1.0$.

The subsamples evaluated include the four magnitude-limited subsamples
in \citeauthor{paperII} ($M \leq 19.3, 19.8, 20.3, 20.8$), with and
without the \citeauthor{paperII} imposed 0.1 mag photometric error limit
on the dataset.  Table 1 shows that the contamination fraction in every
subsample is not a major fraction of the total sample of Carina giant
candidates.  The calculations show the highest level of contamination
($\sim$44\%) is predicted in the subsample of extratidal Carina giants
with no photometric error limits applied and with a magnitude cut of $M
\leq 20.8$; this is expected because this subsample includes the largest
fraction of stars with errors potentially large enough to scatter stars
into the Carina giant region.  In every case, when we apply the actual
\citeauthor{paperII} photometric error limit (i.e., $\sigma_{M},
\sigma_{T_{2}}, \sigma_{DDO51} \leq 0.1$) the expected level of
contamination drops, but not by huge fractions (e.g., for the $M\le20.8$
extratidal sample, from 60 out of 137 to 48 out of 118, or from 44\% to
41\%).  For the entire sample of giant candidates with 0.1 mag error
limits, depending on the magnitude limit, the contamination level is
expected to be about 13\% to 27\%.  These levels of predicted
contamination are clearly not the 100\% proposed by
\citeauthor{MOMNHDF}.

While these calculated estimates of the amount of contamination of our
giant candidate sample due to photometric error may still seem large,
they may also be overestimated.  The calculation that has been performed
is only applicable to the determination of the number of objects that
truly lie outside our giant selection region that we ``incorrectly''
designate as giants.  However, the criteria used to select Carina giants
is not perfect; it was designed to be conservative in its selection of
stars as Carina giants in order to mitigate the amount of contamination,
and in so doing, it purposely excludes other types of expected Carina
members.  It is very likely that many of the objects found just outside
our giant selection region are Carina asymptotic giant branch (AGB)
stars, Carbon stars, horizontal branch (HB) stars, etc.  Thus, objects
found outside the box that scatter into the box are not necessarily
non-Carina members (see, e.g., \S3.3.2).  Moreover, just as non-Carina
stars may be scattered into the giant selection region due to
photometric error, the opposite also occurs: {\it{Carina giants are
    scattering out of our giant selection region due to photometric
    error}}.  The combination of these two effects implies that the
contamination levels reported in Table \ref{contamprob} should be taken
as upper limits.

Finally, we note that the analysis presented here accounts only for
stars misclassified as Carina giants due to photometric error.  It does
not address the issue of how well the boundaries of the
three-dimensional selection region separate Carina giants from other
non-Carina stars that {\it{should}} lie within the giant selection
region (e.g., field halo giants or extreme subdwarfs with colors and
magnitudes that happen to place them along the Carina locus).  On the
basis of the radial velocity (RV) distributions of the Carina giant
sample (\S\ref{sec:spectro} and Figure 4a below) it is possible that
these types of contaminants (which might be {\it assumed} to have
Galactocentric RVs relatively far from $0\,{\rm km\,s^{-1}}$ --- but not
necessarily so; see \S 3.4) make only a small contribution to the total
contamination.  It is worth noting that an accounting of this type of
contamination {\it is} expected from the methodology for subtracting
``background'' described in \citeauthor{paperII}.

\citeauthor{MOMNHDF} claim that the majority of the extratidal giant
candidates identified in \citeauthor{paperII} are misclassified dwarfs
due to (their large adopted) photometric errors, while the results
presented here suggest that only a fraction of these candidate Carina
giants can be misclassified stars.  As shown in Table 2, the true
detection fraction of actual Carina giants predicted by our new
analytical method (column 4) are generally smaller than, but in keeping
with, the general fractions estimated in \citeauthor{paperII} (column
5).  As one more verification that the major source of the difference
between the results of the \citeauthor{MOMNHDF} simulation and our own
calculations here lies in their adopted error distribution, we assigned
each star in our sample an error in the three filters of $\sigma = 0.1$,
the value used in \citeauthor{MOMNHDF}, and repeated our computation of
the expected number of contaminants among the sample of 118 extratidal
giant candidates with $M \leq 20.8$ from \citeauthor{paperII}.  The
result of assuming errors this large is that we should expect $68 \pm 5$
contaminants (compared to the 48 expected contaminants calculated using
the proper error distribution).  This number, when normalized for the
survey area covered by the $M \leq 20.8$ extratidal sample, is $59\pm 5$
deg$^{-2}$.  This is still smaller than, but more consistent with, the
value of 85 deg$^{-2}$ misclassified dwarfs that \citeauthor{MOMNHDF}
estimate our contamination would be {\it were} the error in each filter
for every star as large as 0.1 mag.

\subsubsection{Ranking Giant Candidates}

The probability calculation just described was used to determine the
number of expected contaminants in the \citeauthor{paperII} samples of
giant candidates; however, the calculation also provides a method for
estimating a likelihood that any particular star is a Carina giant (and
the method can of course be generalized to the study of other star
systems).  $P_i$ provides a measure of how closely the $(M - T_2,
M-DDO51, M)_o$ of star $i$ match those of a typical Carina giant star,
taking into account photometric error, and can be used to rank stars
from most likely Carina giants ($P_i = 0$) to least likely Carina giants
($P_i = 1$).  Astrometrists use a similar technique in the proper motion
vector point diagram of star cluster fields \citep[see e.g.,][]{cud85},
with cluster membership probability assigned to each star by comparing
its proper motion and error to that of the cluster mean, and even more
sophisticated joint (trivariate) probabilities that also account for
location of the star in the cluster CMD as well as the spatial location
of the star with respect to the cluster center have been adopted
\citep*[see, e.g.,][]{GJT98}.  A sample of giant candidates selected
using stars in limited ranges of our trivariate photometric membership
probabilities, $P_i$, should be more representative of the true
distribution of Carina giants, and identify the best Carina candidates
for spectroscopic verification.  Once spectroscopy is in hand, one may
also explore probability trends in the contamination level as a function
of different parameters in order to refine future selection criteria.

For example, while in the end \citeauthor{MOMNHDF}, commenting on the
\citeauthor{paperII} Carina survey, admit that ``brighter [giant
candidate] stars, such as the three [{\it sic}\footnote{Twenty-seven
  stars with spectroscopy were discussed in Paper II --- 23 previously
  given in the literature and four with new velocities.  Among these,
  three lay outside the nominal tidal radius.}] observed
spectroscopically, will have smaller photometric errors and will in
general be giants if they are in the giant region'', there remains a
legitimate concern about the level of contamination at fainter
magnitudes.  However, Figure 3, which shows the determined $P_i$ as a
function of $M_o$, suggests that mean calculated probability of being a
contaminant grows relatively slowly with magnitude.  The maintenance of
a more similar error distribution with magnitude is partly a reflection
of \citeauthor{paperII}'s progressive removal of bad data with each
increasedmagnitude limit.

It should be kept in mind that the $P_i$ scale is dependent on the exact
shape of the selection boundary and is only a rank order metric internal
to a particular survey.  Moreover, $P_i$ reflects the probability of
contamination within the adopted selection boundary --- it does not,
strictly speaking, say anything about the true probability of being a
Carina giant.  The degree to which $P_i$ does reflect the probability of
being a giant relies on the degree to which the triavariate selection
boundary genuinely separates Carina giants from non-giants, but this
surface is difficult to know {\it{a priori}}.  From this standpoint,
therefore, the $P_i$ should be looked on only as a guide to the relative
likelihood of membership from star to star.

\section{Spectroscopic Assessment of Carina Giant Membership}
\label{sec:spectro}

\subsection{Observations and Reductions}

In \citeauthor{paperII} we had access to a total of 27 stars in the
Carina field having spectroscopic data from either \citet{mat93} or our
own observations, among which were 21 Carina giants and 6 field dwarfs.
Our photometric selection criteria correctly classified all of them.
Nevertheless, one might be concerned (as suggested by
\citeauthor{MOMNHDF}) that a higher contamination rate in our candidate
sample might be expected at lower Carina densities.  In addition,
\citeauthor{MOMNHDF} considered the three spectroscopically-confirmed
stars in \citeauthor{paperII} that are outside the tidal radius to be
insufficient evidence that we are finding an extended/extratidal Carina
population.

Since \citeauthor{paperII} we have continued spectroscopic follow-up of
our Carina giant candidates, more than tripling the number with spectra
and, as a further test of our procedures, observing many more stars in
the Carina field not selected to be Carina giants.  Here we focus on
what the derived radial velocities (RVs) of these stars tell us about
our photometric selection methodology.  Further consideration of the
{\it dynamics} of the outer parts of the Carina system are presented
elsewhere (R. Mu\~noz et al., in preparation).

Spectra of our Carina giant candidates have been obtained using the
Blanco 4-m + Hydra multifiber spectrograph system at Cerro Tololo
Inter-American Observatory on the nights of UT 2000 March 26-29, 2000
November 10-12 and 2001 October 8-11.  In all runs the region around the
calcium infrared triplet was observed.  For the year 2000 runs, the
Loral 3K$\times$1K CCD and grating KPGLD in first order was used; this
set-up delivers a resolution of $\sim$2600 (or 2.6\AA\ per resolution
element).  For the October 2001 run, we used the SITe 4K$\times$2K CCD
and grating 380 in first order and also inserted a 200 $\mu$m slit plate
after the fibers to improve the resolution to $\sim$7600 (1.2\AA\ per
resolution element).

Our strategy for targeting stars with fibers was to place as first
priority those stars that were selected to be Carina giants.  Remaining
fibers were filled with stars not picked as Carina giant stars as a
means to assess the level of ``missed'' Carina giants with our
photometric selection criteria.  Among these ``fiber filler stars''
observing priority was given to stars that were picked as giants stars
in the 2CD but outside the Carina RGB selection box in the CMD.  Another
category of filler stars were those also lying just outside the 2CD
giant star boundary (for which we eventually obtained good spectra of
six).  Finally, because of the potential that they might be halo HB
stars or HB/anomalous Cepheid stars from Carina, we also targeted stars
bluer than the main sequence turn-off of the field star population in
the CMD.  Any still unused fibers were placed on blank sky positions to
obtain at least a half-dozen background sky spectra.  Four unique Hydra
pointings of Carina targets were obtained in March 2000 and two unique
pointings in November 2000; for these runs 64 fibers were available for
both target and sky observations (fibers not able to be used for any
stars are generally used to collect background sky spectra).  For the
October 2001 run, we obtained two unique Hydra pointings, each observed
on two different nights, with 133 available fibers.  Multiple
observations of the same set-up as well as cross-targeting of individual
stars in the Carina field between different set-ups allows a check on
the RV errors (random and systematic) for individual stellar targets
(Table 3).  To aid the RV calibration, multiple (6 to 13) RV standards
were observed each run, where each ``observation'' of an RV standard
entails sending the light down 7-12 different fibers, yielding many
dozen individual spectra of RV standards.  For wavelength calibration,
the Penray (HeNeArXe) comparison lamp was observed for every fiber
setup.

Preliminary processing of the two-dimensional images of the fiber
spectra was undertaken using standard IRAF techniques as described in
the IRAF {\ttfamily imred.ccdred} documentation.  After completing the
bias subtraction, overscan correction and trimming, the images were
corrected for pixel-to-pixel sensitivity variations and chip cosmetics
by applying ``milky flats'' as described in the CTIO Hydra manual by N.
Suntzeff\footnote{http://www.ctio.noao.edu/spectrographs/hydra/hydra-nickmanual.html}.

Spectral extraction across the PSF of the individual fibers made use of
a local IRAF script much like {\ttfamily imred.hydra.dohydra}, but
developed to simplify Hydra reduction for our RV analysis.  The script
is based on standard IRAF spectroscopic reductions using {\ttfamily
  imred.hydra.apall} for extraction of the fibers, and {\ttfamily
  imred.hydra.identify} and {\ttfamily imred.hydra.reidentify} for
wavelength calibration of the comparison spectra.  The wavelength
solutions are applied using {\ttfamily imred.hydra.refspec} and then
dispersion corrected to a common wavelength range with {\ttfamily
  imred.hydra.dispcor}.  Finally, a master sky spectrum is made and
subtracted from the target stars using the {\ttfamily
  noao.onedspec.skytweak} task.

Our RV reduction is a modified version of the classical
cross-correlation methodology of \citet{TD79}.  However to improve the
velocity precision, we pre-process the spectra by Fourier filtering and,
for the RV template spectra, filtering in the wavelength domain to
remove those parts of the spectra that contribute little to the
cross-correlation template other than noise.  To do so, the filtered
template is multiplied by a mask that is zero everywhere except at a set
of restframe wavelengths of low ionization or low excitation transitions
of elements observed in moderately metal deficient stars.  The latter
process leaves only the most vertical parts of relatively strong
spectral lines to provide the cross-correlation reference.  It is found
that very strong lines that are not on the linear part of the curve of
growth, like the Ca II infrared triplet, are not as useful in this
enterprise, and these lines are actually left out of our
cross-correlation.  Since the output of the correlator is affected only
by the strength of the selected spectral feature compared to that in the
master, this manufactured template spectrum is fairly insensitive to
spectral type.  The Fourier-filtered and masked template, when
correlated with the Fourier-filtered target stars, only responds to
spectral lines that have been selected when they are present.  If a
selected template spectral line is absent in the target star what little
effect is introduced from photon noise is free of bias and detracts
minimally from the correlation.  The foundations and application of this
cross-correlation methodology are described more fully in \citet{sgr2}.

A quality factor \citep[see][]{sgr2} is assessed for each derived
stellar velocity depending on the strength and shape of the correlation
peak with respect to other features in the cross-correlation function,
with $Q=7$ being a solid RV measurement and $Q=4$ the lowest quality
cross-correlation that yields a trustworthy RV (albeit at lower
precision for these typically $S/N \sim 5-6$ spectra than for those with
higher $Q$).  By comparing multiple measures of the same star among the
numerous RV standard observations we have found typical dispersions of
$5\,{\rm km\,s^{-1}}$ for the KPGLD grating setup (year 2000
observations) and $2\,{\rm km\,s^{-1}}$ for the 380 grating setup (year
2001 observations).  Because the spectra of the Carina stars are of
lower $S/N$, we adopt errors twice the above values as representative RV
errors among the higher quality ($Q=6$ or 7) Carina targets.  As we show
below (Tables 3 and 4), these assumed RV errors are not unreasonable.

An unfortunate aspect of this observing program was that it faced
consistently mediocre to poor observing conditions, with each run having
significant clouds and/or poor seeing.  This prevented us from obtaining
good $S/N$ spectra for every targeted star, and generally only the
brightest stars in each fiber set-up had spectra with enough $S/N$ to
derive reliable RVs ($Q > 3$); the results for these stars are presented
in Tables 3, 4 and 5.  As may be seen, the number of successfully
observed Carina targets is substantially less than the many dozens
targeted across the eight different fiber set-ups (generally about 75\%
of the available 64 or 133 total fibers per set-up).  Originally we
sought four to five hour exposures per set-up, but actual integration
times ranged from 1.5 to 3.5 hours because of frequent shutdowns for
weather and technical problems, and these exposures, of course, were
generally compromised with clouds or poor seeing.  The large variation
in integration time is the reason there are few multiply observed stars
from October 2001 (Table 3), even though each fiber set-up was observed
twice on this observing run.

On the other hand, because these spectra have been obtained with fibers
feeding a bench-mounted spectrograph we should not face the same
problems with variable and uncertain entrance aperture illumination and
mechanical flexure that is common to slit spectroscopy; thus we might
not expect to be quite as haunted by systematic RV shifts from observing
run to observing run, or pointing to pointing.  We have noted before
\citep{sgr2} that our initially derived RVs face some systematic offset
from true RVs due to the detailed structure of the manufactured RV
cross-correlation template.  The value of this small offset is derived
by the difference between the measured and literature RV values of the
numerous RV standards observed.  After application of the derived
systematic offsets from run to run, we cannot discern a plausible offset
between the RV systems of those fiber setups that have stars in common
(Table 3).  As one demonstration of this fact, the entries for each star
in Table 3 are listed in RV order, whereas the dates of the observations
do not have any consistency in their relative ordering from star to
star.\footnote{Note that no star in the November 2000 pointings has more
  than one derived RV; however, there is no reason to believe that these
  particular data should behave any differently.}

For stars observed multiple times, the adopted radial velocities (final
entries in Table 5) represent weighted averages, $<RV>$, of the
individual measured velocity values, $v_i$, given in Table 3 for that
star:

\begin{equation}
<RV> = \sum_{i} (\omega_{i}^2 v_i) /  \sum_{i} (\omega_{i}^{2}).
\end{equation}

\noindent The weighting factors take into account the 
fact that the spectra have different $S/N$ while two spectral
resolutions are represented among the data.  For purposes of weighting,
we adopt the system-wide values of $10\,{\rm km\,s^{-1}}$ and $4\,{\rm
  km\,s^{-1}}$ as representative relative velocity errors,
$\epsilon_{i}$ between the lower and higher resolution spectra,
respectively, and obtain weights as

\begin{equation}
\omega_{i} = \omega_{quality, i} \times  (1/ \epsilon_{i}).
\end{equation}

\noindent Previous experience shows that the actual velocity precision is
inversely related to the height of the cross-correlation peak ($CCP$),
but not necessarily in a linear fashion.  Therefore we adopt the
following additional factors in the weighting: $\omega_{\rm quality,i}$ =
0.5/1.0/2.0/3.0 for RVs with $CCP$s in the ranges
($<$0.3)/(0.3-0.5)/(0.5-1.0)/($\geq$1.0).

Table 3 gives the standard deviation, $s(RV)$, of multiply-measured
stars, and reveals that in general the standard deviations are
consistent with the representative RV errors adopted above --- namely 4
and $10\,{\rm km\,s^{-1}}$ errors for the higher and lower resolution
data for spectra with higher quality (i.e., higher $CCP$ and $Q$).
Typically in those cases where $s$ substantially exceeds the above
representative values, at least one of the measures is of lower quality,
as might be expected.  In one case (star C1394) where $s$ is extremely
large one of the two measures is right at our minimum level of
acceptable quality, and this quality is evidently overestimated.

A comparison of our Hydra RVs to previously published values for stars
in common with \citet{mat93} and \citeauthor{paperII} is given in Table
4.  For the stars C1547 and C2282 we obtain RVs within a few ${\rm
  km\,s^{-1}}$ of the \citeauthor{mat93} values.  For star C2774 our RV
is $17\,{\rm km\,s^{-1}}$ different than that of \citeauthor{mat93}, but
this Hydra RV also has a relatively low $CCP$ and $Q$, so the difference
is not surprising.  We have also remeasured with Hydra the four stars
with du Pont telescope spectra (having typical RV errors of $10-15\,{\rm
  km\,s^{-1}}$) presented in \citeauthor{paperII}.  The new spectra are
both of higher resolution and better $S/N$, so here the comparison
provides a less useful quantitative evaluation of the new data.
However, it is interesting to note that for two of the du Pont-observed
stars (C2501583 and C2103156) the new Hydra RVs are much closer to the
canonical Carina RV.  The other two \citeauthor{paperII} stars have
consistent RVs between du Pont and Hydra.

Table 5 summarizes the Hydra results for all stars in the Carina field
having spectra with high enough Hydra $S/N$ to derive reliable RVs, or
that have been published before by \citet{mat93}.  The entries include
the star name, fiber coordinates, date observed (for Hydra
observations), photometric data (dereddened values), and the derived
heliocentric RV ($V_{\rm r}$), $CCP$ and RV quality ($Q$) values for each star
(we denote RVs derived from \citeauthor{mat93} with a $Q=M$).  The
column ``Cand'' in Table 5 specifies whether the star was originally
selected to be a Carina giant (``Cand'' = ``Y'') or not (``Cand'' =
``N'').  The Table 5 sample of 134 stars includes 74 stars selected as
candidate Carina giants using the \citeauthor{paperII} criteria and 60
``fiber filler'' stars including six known dwarf stars from
\citet{mat93}.  The final column in Table 5, ``Mem?'', summarizes our
evaluation of the true membership to the Carina system according to the
procedures outlined below.

\subsection{Defining Carina RV Membership}

Armed with new RV data, we must determine a method by which to judge
what is a Carina member.  Fortunately, at Carina's position in the
Galaxy (near the direction of anti-rotation, $[l,b] =
[260.1,-22.2]^{\circ}$) and systemic heliocentric velocity ($V_{\rm r} =
223.1\,{\rm km\,s^{-1}}$; \citealt{mat98}), the random non-Carina stars
in the field should be dominated by stars having substantially different
RVs than Carina.  Among the stars selected to be giant candidates, the
primary expected contaminants will be: (1) stars errantly selected due
to photometric errors, which are presumably dominated by the more
populous foreground disk dwarfs and which will be obvious by their near
zero RVs, (2) metal-poor dwarfs with low magnesium abundances, which at
the survey magnitudes will be dominated by Galactic thick disk stars,
and which will (given the asymmetric drift of this Galactic population)
have RVs somewhere between that of thin disk stars and Carina, but
closer to the former, and (3) random halo giants not related to Carina.
If the Galactic halo is randomly mixed and has close to zero net
rotation, the mean halo giant velocity in the Carina direction will, in
fact, have a heliocentric RV close to that of Carina, but the broad
velocity dispersion of a dynamically hot halo \citep*[$\sim
100-150\,{\rm km\,s^{-1}}$; e.g.,][]{NBP85, CL86, lay96, sir04} means
that only a fraction of these stars will have RVs lying within the
tighter range of Carina stars.  On the other hand, if the halo is
instead not well-mixed and networked with dynamically cold
substructures, there is the possibility of both large density variations
in ``field giant'' density as well as the potential for other
substructure in the field with any particular mean velocity (e.g., see
\S3.4), including, in principle, one near that of Carina.  But overall,
from these general arguments, we conclude that RVs should provide a
fairly reliable means to discriminate Carina members from non-Carina
stars in the same field.
 
Figure 4 gives RV histograms of all 134 stars in Table 5.  The
distributions of stars selected photometrically as Carina giant
candidates (Fig.\ 4a) and those not selected to be Carina giant
candidates (Fig.\ 4b) are shown separately.  The clear signal of Carina
stars near its systemic velocity of $\sim223\,{\rm km\,s^{-1}}$ is
obvious in Figure 4a; in general these Carina stars are fairly separated
from the small fraction of stars photometrically selected to be ``Carina
giants'' but having RVs inconsistent with Carina membership (i.e.,
\citeauthor{paperII} ``false positives'').  The near zero heliocentric
velocities for most of the latter stars suggests that the primary source
of the small amount (quantified below) of contamination is the Galactic
thin disk, with stars presumably making it into our sample through
photometric error (however, see \S3.4).  Despite the rather clear
delineation of Carina stars in the top panel of Figure 4, we seek a
``fair'' way to discriminate members versus non-members because: (1) A
few stars selected as Carina giants have more intermediate RVs, which
gives them ``borderline'' Carina membership depending on the criteria
selected.  (2) We are also interested in possible Carina membership
among the ``filler star'' sample (Fig.\ 4b), which spans a larger RV
range.  Among the latter, there is a peak in the RV distribution at the
Carina systemic RV, indicating the presence of a fair number of Carina
stars.  We can use the appearance of the Carina RV distribution in the
top panel to guide how Carina members among the filler stars might be
identified.

To select stars as Carina RV members objectively, we determine the mean
value of the ``Carina peak'' in Figure 4a using an iterative rejection
of 2.5$\sigma$ outliers.  In terms of the number of $\sigma$, this is
more restrictive than the $3\sigma$ limit utilized by, for example,
\citet{wil04} in their study of the Ursa Minor and Draco dSphs, but in
terms of absolute velocities it is larger because the latter work had
better, $2.4-2.9\,{\rm km\,s^{-1}}$ RV precision.  However, while we are
forced to accept a broader absolute RV range to accommodate our larger
intrinsic RV errors, we can ``afford'' to do this because we have vetted
our stars to be photometrically-selected giants and a comparison of
Figure 4a to Figure 4b shows that this step must be at least partly
effective in lowering contamination.  The primary expected contaminant
with a near-Carina RV would be a halo giant star, but any giant star
slipping through our photometric selection criteria with a similar RV
and distance (i.e., position in the CMD) as Carina RGB stars are most
likely to {\it be} Carina RGB stars.\footnote{A similar logic was
  applied in the assessment of RR Lyrae stars around the Sculptor dSph
  by \citet{IP79}.}  Upon convergence of the iterative procedures, those
stars in Figure 4a lying within 2.5$\sigma$ of the mean, where $\sigma$
is found to be $16.6\,{\rm km\,s^{-1}}$, are kept as Carina members.
Sixty-one stars remain in the converged fit, and yield a mean Carina RV
of $222.8\,{\rm km\,s^{-1}}$, only $0.3\,{\rm km\,s^{-1}}$ from that
found by \citet{mat93} --- but note that all 17 of the latter stars are
included among the former.  For the ``filler star'' sample, we adopt
these same RV limits {\it as a starting point} to hunt for additional
Carina members, but also explore the positions of these stars in the CMD
as an additional criterion to judge for likely membership to Carina
(\S3.3.2 and \S3.3.3).

\subsection{Evaluating the \citeauthor{paperII} Carina Giant Selection}

\subsubsection{Membership Rates and False Positives}

The subsample of stars observed with Hydra were selected for maximal
efficiency in fiber usage with the multi-object spectrograph, and among
the stars observed there is uneven $S/N$ due to variable weather
conditions, integration times and stellar fluxes.  Therefore, this
subsample is not easily described in terms of statistical completeness
with regard to sky position, color, magnitude or photometric errors.  As
a test of our photometric survey the most straightforward assumption is
that the RV sample represents a reasonable proxy of the the full Carina
giant candidate list so that we may compare our expectations for the
contamination level with the ``false positive'' fraction number among
the 74 Carina giant candidates having measured RVs.

Based on the definition of Carina RV membership adopted in \S3.2, we
derive the fractional number of ``true members'' for various subsamples
drawn from the Carina giant candidate sample as shown in Table 2.  The
{\it spectroscopically-verified} membership fraction numbers, given by
the fraction of stars in Table 5 with ``Cand''=``Y'' that also have
``Mem?''=``Y'', are compared to both the member fractions rates
predicted from the analysis in \S2 and to the background rates derived
in Paper II.  As may be seen, the actual observed rates of true Carina
members among the Carina giant candidates are similar to those
previously predicted here and in \citeauthor{paperII}, and, obviously,
there is nowhere near the ``100\% contamination'' suggested by
\citeauthor{MOMNHDF}.  Rather, the spectroscopic membership fractions
are close to, and more often than not {\it higher than}, those predicted
from our analysis in \S2.

Overall we find that our $M,T_2,DDO51$ methodology is rather efficient
at identifying true Carina members across the various subsamples listed
(44-94\% members, depending on the actual magnitude, magnitude error and
spatial location limits imposed on the \citeauthor{paperII} photometric
database).  The trend is for lower membership fractions among the
deeper samples, presumably because of a larger fraction of stars with
larger photometric errors and because the giant branch here is bluer and
closer to the color of the bulk of the contaminants.  Across our full
survey to $M=20.8$ the spectroscopic membership rate is 81\%.  In
comparison, for example, the \citet{mat93} selection of Carina
candidates proved only 74\% efficient in finding true members, and this
was among a sample of the very brightest ($M<18.3$, or a magnitude limit
$\sim1.0$ mag brighter than our brightest magnitude limit), reddest
(i.e., most obvious compared to the field star population) candidates in
the Carina {\it core} where the density of members is several orders of
magnitude higher than in much of the area we have explored here.

The overall high membership rate among the Carina giant candidates is
similar to --- though less than --- the positive results we have had
with other, similar Washington$+DDO51$ selected giant candidate samples
we have studied spectroscopically in this series of papers: (1) In an
analysis of the Ursa Minor dSph by \citet{pal03} identical to the
\citeauthor{paperII} analysis of Carina, our technique is 100\% accurate
in classifying 84 giant stars, 60 dwarf stars, and even one field giant,
among 154 spectra obtained by \citet{har94} and \citet*{arm95}.  The
remaining 9 stars are found to be RV members of UMin, but they were not
selected photometrically as candidate stars because the majority appear
to be AGB stars, which are found just outside the adopted CMD limits of
the \citeauthor{pal03} RGB locus.  (2) In the similar survey of Leo I by
\citet{soh03} and S. Sohn et al.\ (in preparation), we have obtained
spectroscopy of 85 Leo I giant candidates from the center and out to 1.3
times the King limiting radius and have verified 100\% of them to be Leo
I giants.  (3) In a study of the Sculptor system \citep{wes05}, we have
found 97\% of the 146 photometrically-selected Sculptor giants to be RV
members of the dSph, including stars to 1.5 the King limiting radius.
(4) As part of the program by \cite{guh04} Keck spectra for 30 Andromeda
I giant candidates and 24 Andromeda III candidates identified with the
same techniques have been obtained (J. Ostheimer et al., in
preparation).  All of the And I/III giant candidates are found to be RV
members of these systems, apart from only two And I candidates that are
M31 halo giants (which share the same RGB and distance as And I, and can
hardly be considered as ``failures'' of our method).  These results for
other dSphs are additional evidence that our photometric selection
method works well and that it is a most efficient way to find the
``needle-in-the-haystack'' giant stars needed for study of true dSph
stars well outside of the core radius.

It is interesting to assess the primary source of the false positive
detections.  From the discussions in \citeauthor{paperII},
\citeauthor{MOMNHDF}, and \S2, and the distribution of RVs in Figure 4a
(see \S3.2), the expectation is that the primary source of contaminants
are stars whose photometric errors are sufficient to scatter them into
our selection criteria.  Figures 5a and 5b show the location of the
false positives (open circles) in the CMD and 2CD respectively.  As may
be seen, many of the stars are near at least one of the selection
boundaries, but some are not, and require larger photometric errors to
scatter them into our sample.  The left panels of Figure 6 demonstrate
that, indeed, the false positives are among those stars with larger
photometric errors and, as expected, those with the highest calculated
probabilities of being a contaminant.  Indeed, near 100\% reliability
would have been found had we limited our spectroscopy to stars having
photometry of a precision near the median magnitude errors ($\sigma \sim
0.035$ mag) of the \citeauthor{paperII} survey.  On the other hand, such
a limitation would have {\it missed} a number of the actual Carina
members found in the spectroscopic sample (about half of the stars
selected with the poorer photometry are found to be Carina members),
including most of the extratidal examples.  The main conclusion to be
drawn from the left panels of Figure 6 is that when better photometry is
available the $M, T_2, DDO51$ technique as practiced in
\citeauthor{paperII} works even better\footnote{This statement is borne
  out by the near 100\% identification success for the $M, T_2, DDO51$
  surveys of the Ursa Minor, Sculptor and Leo I dSphs discussed in the
  previous paragraph; all three of these surveys have generally higher
  quality photometry than the present Carina study.}; such photometry is
now available for Carina (R. Mu\~noz et al., in preparation).  However,
even with modest photometric quality, reasonable membership
identification rates have been achieved.

Figure 7 shows the distribution of RVs as a function of $P_i$, and
demonstrates the usefulness of the $P_i$ for identifying those stars
most likely to be Carina giants ($P_i = 0$).  As $P_i$ increases, its
discriminatory power declines, but with improved photometry and improved
selection boundaries (see \S 3.3.2) the $P_i$ discrimination of Carina
giants should improve.

\indent From the standpoint of claims for extratidal giant star
densities, Table 2 suggests that \citeauthor{paperII} has, in fact,
underestimated its background levels.  In the {\it worst case} scenario
--- i.e., for our deepest samples to $M=20.8$ where the level of
contamination is predicted and observed to be highest --- Table 2
suggests that the true densities of Carina stars may actually be as low
as 82\% of the value calculated in \citeauthor{paperII} across our
entire survey area or 64\% of the value calculated in
\citeauthor{paperII} for the ``extratidal'' regions where the results
are most contentious.\footnote{In the best case scenario --- i.e., for
  our $M \le 19.3$ sample -- we are actually finding true ``extratidal''
  members at a rate {\it higher} than predicted in \citeauthor{paperII},
  but this sample includes only seven stars.  }  Nevertheless, as
discussed in \S1.2 and as is clear from Figure 1, raising the estimated,
subtracted extratidal backgrounds accordingly (e.g., by a factor of two
in the worst, $M=20.8$ extratidal case) will {\it not} erase the
presence of an excess density of stars outside the nominal tidal radius
of Carina.  Moreover, as we now demonstrate, the existence of additional
true Carina giants {\it missed} by our \citeauthor{paperII} selection
criteria means the true density of such stars is actually higher than
the Table 2 numbers would imply.

\subsubsection{``Missed'' Carina Members}

We have shown that among stars selected in \citeauthor{paperII} to be
Carina giants, a major fraction of them are indeed true Carina members.
But if we are interested in evaluating the true densities of Carina
giants at any given Carina radius, it is useful to understand how {\it
  complete} was the Carina star selection in \citeauthor{paperII}.  A
lower limit to the ``missed'' density of Carina giants may be derived
from the numbers of plausible RV giant members found among our ``fiber
filler'' sample.  The starting points for this evaluation are the
potential ``missed'' Carina RV members shown by the shaded region in
Figure 4b.

Among these, we focus first on the stars observed spectroscopically from
the ``fiber filler'' category of stars selected to be giant stars, but
that lie outside the CMD selection for {\it Carina} giants.  Twenty-nine
of these stars were successfully observed, and, by the RV membership
criteria of \S3.2, 12 of these stars have RVs in the ``Carina-member''
range.  The CMD and 2CD distributions of these stars are shown in
Figures 5a and 5b by the filled circles redward of the field star main
sequence turn off (near $[M-T_2]_o=0.75$).  Interestingly, all but two
of these stars lie quite near the edge of the CMD selection boundary.
The star with Carina-like RV near $M\sim14.8$ (C2300060) is too bright
to be part of the Carina system while the red star near $M\sim16.8$
(C2501002) could be part of the Carina system only as some kind of
post-asymptotic giant branch (PAGB) species (see \S3.3.3).  Of the
remaining ten stars with Carina-like RVs three appear to be at the
Carina AGB tip, while others fall just above and below the CMD selection
box.  Because of the proximity of these stars to the Carina RGB {\it
  and} their Carina-like RVs, these stars are almost certainly true
Carina members.  Inspection of the photometric errors associated with
the red ``fiber filler'' stars (Figure 6, right hand side) indeed shows
that {\it some} of these stars may have been ``missed'' by the
\citeauthor{paperII} Carina selection criterion because photometric
errors scattered these stars {\it out of} our sample.  On the other
hand, some of these stars seem to have relatively good photometric
uncertainties, so that it is likely they have been missed because the
\citeauthor{paperII} CMD selection criteria were too conservative.  It
is found that had the CMD selection limits been extended in both
directions of luminosity by 0.15 mag, seven of these stars would have
been selected as members, thereby increasing the completeness of the
sample {\it with virtually no change in the spectroscopic membership
  fractions in Table 2}.\footnote{When the CMD limit is expanded in this
  way, all of the fractional membership rates actually {\it increase} by
  0-3\%.}  In the Table 5 summary, we have marked as Carina members all
red ``fiber filler'' candidates with Carina-like RVs and lying just
outside the Carina RGB boundary in the CMD.  This includes the three
stars with Carina RVs above the tip of the boundary in the CMD.

A secondary filler star category is red stars just outside the selection
criteria in {\it both} the 2CD and the CMD.  Among the six of these
stars observed, none are found to have a Carina-like RV.  A seventh star
falling inside the CMD selection but just outside the 2CD selection is
also found not to be a Carina RV member.  These results suggest that the
color-color selection criterion adopted in \citeauthor{paperII} was
reasonably placed for discriminating that part of parameter space
well-populated with Carina giants from that not well-populated.

\subsubsection{Additional Carina Members}

Another category of fiber filler stars targeted spectroscopically were
blue stars.  Although these do not bear directly on the efficiency of
our {\it giant star} selection, it is of interest to know whether
additional Carina members lie among the blue stars in the Carina field.
As shown in Figure 5, eighteen blue [$(M-T_2)_o < 0.75$] stars have been
targeted; seven of these blue stars (those with filled symbols in Figure
5a and 5b) have RVs that we have defined as Carina-like according to the
criterion in \S3.2.  Two of the blue stars near $M \sim 18.7$ (C4156 and
C2563) lie in the region of the Carina field CMD occupied by Carina
anomalous Cepheids (Figure 8) in the survey by \citet{ora03}, and these
two stars have been identified as variable by these authors.  Given that
these two stars also lie within the Carina King limiting radius, they
are almost certainly Carina members.

Recognizing that pulsating variable stars have RVs that shift tens of
${\rm km\,s^{-1}}$ from their mean velocity it is worth investigating
whether there may be other Carina pulsators with RVs slightly outside
the RV membership limits adopted in \S3.2.  Figure 9 shows the
distribution of RVs for all eighteen targeted stars with $(M-T_2)_o <
0.75$.  The distribution is more or less bimodal with one group having
$V_{\rm r}$ closer to zero.  Among the other nine stars, seven are tightly
clumped at the nominal Carina RV, but two lie within $15\,{\rm
  km\,s^{-1}}$ of the lower RV-membership limit from \S3.2.  One of
these stars (C3994) lies among the anomalous Cepheids in the CMD and the
other (C3001272) lies on the Carina HB near the instability strip (both
stars are marked by squares in Fig.\ 8); both lie well inside the Carina
King limiting radius and we consider them both to be Carina members
having RVs modulated by pulsation.

On the other hand, the four brightest blue stars with Carina-like RVs
are too bright to be Carina anomalous Cepheids and are 4-5 magnitudes
brighter than the Carina HB.  Yet it is somewhat remarkable that these
four stars should have such similar RVs to each other and to Carina.
While it is plausible that these stars represent an unrelated moving
group of halo HB stars that just happen to have the same RV as Carina,
the two faintest of the four bright blue stars (C1211401 and C3897) lie
within 0.7 tidal radii of the center of Carina, which represents a small
fraction ($\sim$ 10\%) of the total survey area.  These two stars are at
reasonable magnitudes to be post-AGB (PAGB) stars from Carina's $\sim2$
Gyr population (e.g., see PAGB evolutionary tracks from \citet{blo95},
which show that for Carina stars no younger than $\sim2$ Gyr the PAGB
can have $-4.5 \lesssim M_V \lesssim -2.8$, or about 2.4-4.1 magnitudes
above the Carina HB).  This PAGB ``sequence'' might even extend to the
redder star (C2501002, also shown in Figure 8) at the same magnitude and
that also possesses a Carina RV.  On the other hand, the two brightest
blue stars with Carina RVs (C3509707 and C2502058) would seem to require
more massive, even younger progenitors for explanation as Carina PAGB
stars.  Recently the existence of Carina stars as young as 0.6 Gyr or
younger have been reported by \citet*{HGV00} and \citet{mon03}.  Such
stars may be sufficiently massive to create PAGB stars that approach the
brightness of stars C3509707 and C2502058 \citep{blo95}.  In addition,
were the latter stars some variety of pulsational variable, it is
possible that they are ``caught'' near maximum brightness.  In this
regard it is interesting to note that one of the Carina field variables
identified by \citet{ora03} has the same apparent magnitude as the two
brightest blue RV stars (see Fig.\ 8), although these authors actually
identify this star as a 0.3 day period RR Lyrae. One other problem with
the possible association of C3509707 and C2502058 to a $<1$ Gyr Carina
population is that those young Carina stars are preferentially
concentrated to the core of the dSph, whereas the two bright blue stars
with Carina RVs are well outside the King limiting radius (see the two
open squares with Carina RV in Fig.\ 10a).  Alternative explanations
that could accommodate scenarios with older populations might include
that the two bright stars come from older AGB progenitors that endured
less mass loss, or that they are ``born-again AGB stars
\citep[e.g.,][]{iben83}, although finding {\it two} examples of stars in
such a short phase of stellar evolution would seem extremely unlikely.
In the end, we regard the status of these two stars in our sample as
still very uncertain.  It will be interesting to obtain RVs of the other
bright blue stars in the field to see if additional stars at the same
magnitudes have Carina RVs, a situation that would be even harder to
explain away as mere field contamination.

Finally, the bluest star for which we have obtained a spectrum (C1201)
also seems to have a Carina-like RV.  This star lies in a CMD position
that is not inconceivable for a fading PAGB star.

In the end, it is plausible that at least seven of the nine blue stars
with RVs similar to that of Carina are true Carina members (using the
expanded RV acceptance range discussed above); the fainter five are
marked as such in Table 5, whereas the two near $M_o=16.8$ are marked as
``Mem?''=``Y?''.  The remaining two stars might be extremely bright
Carain PAGB stars, but this is unlikely because (1) their origin is
challenging to explain, and (2) their position outside the nominal
Carina tidal radius is incompatible with the least unlikely scenario
involving an association with rather young progenitors.

\subsection{A Second Look at the RV Distribution}

To this point we have regarded our targeted Carina giant stars (Fig.\ 
4a) having $V_{\rm r} \lesssim 100\,{\rm km\,s^{-1}}$ as being most likely
dwarf star contamination from the Milky Way disk (and in \S3.3.1, along
with Fig.\ 6 we argued that larger photometric error among the false
positives points in this direction).  However, another possibility is
that these stars represent giant stars from other halo substructure
having these velocities.  We arrive at this notion via the results of a
large area, deep survey for giant stars we have conducted
\citep[see][]{maj99,maj04} around the Magellanic Clouds and including
fields within $5-10^{\circ}$ of Carina.  The RV distribution of the
giant stars in these fields is nearly bimodal, with one large group of
stars concentrated with RVs between $100-200\,{\rm km\,s^{-1}}$ less
than that of Carina \citep[e.g., see right side of Fig.\ 6 in][]{maj04}.
This RV clustering of stars, which have projected giant star distances
ranging to many tens of kiloparsecs, is already observed to span tens of
degrees on the sky and it is not inconceivable to find this same
population of stars in the relatively nearby Carina field.  While in
both surveys the ``giant candidates'' have been found using similar $M,
T_2, DDO51$ photometric techniques, the ``giants'' in the low RV group
in these Magellanic periphery fields include stars that are redder and
brighter and are therefore quite reliable identifications --- i.e. in
{\it that} survey the low velocity giant candidates are {\it not} likely
to be dwarfs scattered into the survey due to photometric errors.

It is also curious that we find a similar ``Carina giant'' RV
distribution in Figure 4a to that of the blue star sample in Figure 9.
That there are fainter blue stars with RVs at similar, low $V_{\rm r}$ is
consistent with the possibility that both are tracing one halo
substructure (white dwarfs are the only likely nearby, disk type star at
these Galactic coordinate and with colors this blue and magnitudes this
faint, but their density is less than what we observe for blue stars
here).  However, in this case, the relevant blue stars would have to be
PAGB or anomalous Cepheids to be counterparts to giants at Carina-like
magnitudes.

While in the end if these false positive stars are field giants and not
dwarfs they still represent {\it Carina} giant contaminants, but we
argue that this is less of a failing for our search methodology in the
sense that the strategy is meant primarily to eliminate dwarf star
contaminants.  If there happen to be other, non-Carina giant stars at
about the distance of Carina with similar metallicity (i.e., located at
a similar place as the Carina RGB locus in the CMD) these cannot be
distinguished from Carina giants with the Washington$+DDO51$ strategy we
have used.  It should be noted, however, that the background subtraction
method utilized in \citeauthor{paperII} is intended to accommodate this,
along with other kinds of, contamination.

Yet another possibility is that the ``false positive'' stars have
``correct'' photometry --- i.e., they are not {\it scattered} into our
selection sample --- but are actually weak-lined dwarf stars.  As shown
in Paper I, this requires extreme, [Fe/H] $<-2.5$ or so metallicities.
We cannot presently discount this possibility, and the numbers of these
stars are not wholly inconsistent with the projected numbers of stars
given in MOMNHDF.

\subsection{Carina Members Well Past the King Limiting Radius}

Table 5 summarizes those stars observed spectroscopically that we
consider to be Carina members after consideration of both their RVs and
their position in the CMD and 2CD.  In Figure 10a we show the
distribution of all available RVs measured to date for stars in the
Carina field as a function of their elliptical radius from the center of
the galaxy.  The elliptical radius for a star, $r_{\rm e}$, is defined to be
the semi-major axis of the ellipse on which each star lies that has
Carina's center and ellipticity (from \citeauthor{IH95}).  We normalize
$r_{\rm e}$ such that stars within the \citeauthor{IH95} tidal radius have
$r_{\rm e} < 1$, while those outside have $r_{\rm e} > 1$.  Stars selected in
\citeauthor{paperII} to be Carina giants, stars selected to be
interesting blue stars, and all other stars observed are shown as
circles, squares, and triangles respectively, with solid symbols used
for those stars denoted as Carina members in Table 5 and open symbols
for non-members.

A significant new result of the present work is the verification now of
a total of 13 Carina members outside of the nominal Carina King limiting
radius.  This includes stars to $r_{\rm e}=1.44$, which is well outside the
errors in the determination of the location of that radius (see Fig.\ 1
profiles, for example).  The existence of true Carina members in the
Carina radial profile ``break population'' found in \citeauthor{paperII}
is beyond doubt.  Based on the RV-members identified in this paper, we
also conclude that the density of the break population is within 36\% of
that measured in \citeauthor{paperII}, even ignoring the ``missed''
Carina giants (\S3.3.2) due to the conservative CMD selection criterion
adopted in that study.

Figure 10b shows the sky positions of the Carina members in Figure 10a
using the same symbols.  The paucity of members in the outer parts of
the photometrically surveyed region is partly a reflection of the
overall sampling: The placement of our Hydra pointings has tended to
favor radii closer to the King limiting radius in order to make optimal
use of the multifiber capability.  However, it is interesting that the
two Hydra set-ups that probed to large elliptical radii lie along the
Carina {\it minor axis}, and very few Carina members are found at large
radii in those directions.\footnote{It is perhaps significant that the
  Carina giant candidates probed to the southeast --- among which there
  are no Carina members found --- are also in the part of the
  \citeauthor{paperII} survey derived from CCD imaging taken in poorer,
  non-photometric conditions.}  Indeed, although this apparent trend
must be tempered by the irregular field sampling, the majority of the
Carina members seem to be concentrated towards the major axis of Carina.
This may be an indication that the break population is indeed tracing
tidal debris rather than an extended Carina halo.

\section{Conclusions and Summary}

This paper is an update and review of ongoing studies of the structure
of the Carina dSph.  \S1 summarized and analyzed the results of previous
photometric surveys of Carina and attempted to resolve the disparity in
claims with regard to its extended structure via a focus on the relative
``signal-to-backgrounds'' of these surveys.  Understanding backgrounds is
key to proper assessment of low density structures, and attempts to {\it
  minimize} background through photometric filtering (e.g., as done in
Paper II) are shown to yield orders of magnitude gains in contrast for
the outer parts of the dSph.  Similar comparisons and arguments are made
relative to a parallel study of the Sculptor dSph by \citet{wes05}.

In \S2 we looked at statistical analyses of the residual
contamination/background level expected in photometrically-filtered dSph
studies like that of the Paper II survey of Carina.  We first assessed
the simulations of the Paper II contamination level by MOMNHDF and
pointed out several problems with their analysis, most notably: (1)
MOMNHDF have insufficiently modeled the overall method of sample
selection in Paper II (which translates, e.g., to an incorrectly
characterized color range of the Paper II Carina giant candidates),
because their simulations of propagated photometric error focus only on
effects within the $(M-T_2, M-DDO51)$ color plane and ignore the
mitigating effects of the equally important Paper II selection of giant
candidates in the $(M-T_2,M)$ color-magnitude plane; and (2) MOMNHDF
substantially overestimated --- by a factor of {\it three} --- the
typical photometric errors of stars in Paper II.  When the latter
problem {\it alone} is corrected for in the MOMNHDF analysis by use of
the {\it proper} Paper II error distribution, an estimated Carina
contamination level is obtained that is similar to that previously
estimated in Paper II.  Thus, the MOMNHDF suggestion that the
\citeauthor{paperII} finding of a break population in the Carina radial
density profile claimed was artificially produced by an order of
magnitude underestimation of background levels is clearly incorrect.

\S2 also provides a new, alternative and independent statistical
analysis of the \citeauthor{paperII} Carina giant candidate sample using
a Bayesian methodology and derives an estimated contamination level for
that data set that is only slightly higher than that originally derived
in \citeauthor{paperII}.  Thus, both the MOMNHDF and our own {\it a
  posteriori} analyses --- when properly matched to the Paper II
methodology and error distribution --- support the general finding in
\citeauthor{paperII} that Carina has a prominent break population
extending to at least several times the King limiting radius.  Such an
extended Carina population was also previously found by \citet{KSH96},
and the work of \citeauthor{IH95} and \citet{mon04} also favor its
existence.  The {\it a posteriori} analysis of photometric errors given
here also provides a prescription for how to rank order
photometrically-selected samples of stars for the likelihood that they
lie within pre-specified regions of multivariate color-magnitude space
(for present purposes applied to regions in the combination of
$M,T_2,DDO51$ space where Carina giants lie).

We agree with the sentiments expressed {\it both} in
\citeauthor{paperII} and \citeauthor{MOMNHDF} that spectroscopic
confirmation provides an important check on the veracity of the
\citeauthor{paperII}, or any photometrically selected, dSph giant star
sample.  On the other hand, we disagree with the sentiments of
\citeauthor{MOMNHDF} that {\it any} scientific results are precluded
before a complete set of spectroscopic data are in hand.  After all, the
majority of published surveys of the structure of Galactic dwarf
spheroidal galaxies have been based on simple starcounts, often from
photographic data, and with {\it no} attendant spectroscopy and {\it no}
means to separate likely foreground/background stars from likely dwarf
spheroidal member stars, information that vastly improves the
signal-to-background of a survey.  As long as the case can be made that
the \citeauthor{paperII} giant star candidates have been selected in an
unbiased way {\it and} that the background has been properly accounted
for, then the statistics of ``likely'' Carina giant stars should (1)
yield plausible information about the shape of its stellar distribution
and (2) give results that are a significant improvement over studies to
the same magnitude limit that have no (or less complete) means to lower
the background contribution.

The statistical analyses given in \S2 already make this case.  However,
in \S3 we presented new spectroscopic data to extend the subsample of
RVs derived for stars in the Carina field, and these data lend further
evidence that the \citeauthor{paperII} Carina giant candidate background
assessment was reasonable and that the radial profile derived there
provides a reasonable description of the true Carina profile.  Most
importantly, we have proven the existence of a significant population of
Carina stars beyond the nominal King limiting radius (with now thirteen
RV members there), and stretching to 1.44 times beyond that radius.
While the Washington$+DDO51$ method proves to be an efficient means to
identify actual Carina giant stars, it is found that the actual CMD
criterion used in \citeauthor{paperII} may have been too conservative,
offsetting the somewhat underestimated background in
\citeauthor{paperII}, so that the actual density of Carina giant stars
outside the King limiting radius may not be far from the actual levels
predicted in that paper.

An unexpected and interestiqng additional finding from the spectroscopic
program is that among stars having the same RV as Carina we have
identified potential members of its PAGB population.  If other stars at
a similar position in the CMD turn out to have Carina-like RVs it would
suggest that the PAGB is prominently represented in Carina.

Proof that an extended, break population of stars in Carina exists does
not resolve the question of what exactly this break population {\it is}
--- i.e., whether it is a bound or unbound component of Carina.  The
actual spatial distribution of RV members in Figure 10b is somewhat
suggestive of tidal tails, but further sampling of the
\citeauthor{paperII} candidate list is clearly needed to verify this
trend.  On the other hand, the similarity in radial profiles between
Carina and the Sagittarius dwarf spheroidal \citep{sgr1} --- for which
the break population is established to arise from prominent tidal tails
--- is compelling support for the view that Carina may also be
undergoing substantial tidal disruption.  Further evidence in this
direction comes from a completely new, deeper, higher quality, and more
extensive Washington$+DDO51$ photometric survey by R. Mu\~noz et al.\ (in
preparation) which essentially confirms the existence of a prominent
break population in Carina.  Additional spectroscopy of candidates from
this new survey, presented in a future contribution, demonstrates that
(1) the Carina break population extends even farther in radius, and (2)
the dynamics of these stars is consistent with the idea that they are
tidal debris.

\acknowledgements

We appreciate useful conversations with David Nidever, Robert Rood,
Michael Siegel and Verne Smith.  This material is based upon work
supported by the National Science Foundation under Grant Nos.\ 9702521
and 0307851, a Cottrell Scholar Award from The Research Corporation, the
David and Lucile Packard Foundation, NASA/JPL contract contract 1228235
and The F. H. Levinson Fund of the Peninsula Community Foundation.  PMF
and RRM hold Levinson Graduate Fellowships awarded through the Peninsula
Community Foundation.  PMF is supported by the Virginia Space Grant
Consortium.  DG gratefully acknowledges support from the Chilean {\sl
  Centro de Astrof\'\i sica} FONDAP No.\ 15010003.

\begin{deluxetable}{llccccc}
  \tablewidth{425pt} 
\tabletypesize{\small} 

\tablecaption{Expected Number of Contaminating Stars Found in the Giant
  Selection Region of $(M - T_2, M - DDO51, M)_o$ Space.}

\tablehead{ \colhead{Sample name\tablenotemark{a}} &
\colhead{Error Cut\tablenotemark{b}} &
\colhead{$M$ Limit}& 
\colhead{$N_{\rm stars}$} &
\colhead{$\langle N_{c}\rangle$\tablenotemark{c}} &
\colhead{$\sigma_{\langle N_c \rangle}$}\hspace{-5pt}\tablenotemark{d} & 
\colhead{$\%contam$\tablenotemark{e}}
}

\startdata

All giants & none & $\leq 20.8$ &  868 & 251 & 6 & 29  \\
All giants & $\sigma_{M}$, $\sigma_{T_2}$, $\sigma_{DDO51} \leq 0.1$ &
$\leq 20.8$ & 802 & 214 & 5 & 27   \\
All giants & none & $\leq 20.3$ & 592 & 142 & 4 & 24  \\
All giants & $\sigma_{M}$, $\sigma_{T_2}$, $\sigma_{DDO51} \leq 0.1$ &
$\leq 20.3$ & 552 & 117 & 4 & 21   \\
All giants & none & $\leq 19.8$ & 395 & \phn79 & 3 & 20  \\
All giants & $\sigma_{M}$, $\sigma_{T_2}$, $\sigma_{DDO51} \leq 0.1$ &
$\leq 19.8$ & 385 & \phn72 & 3 & 19 \\
All giants & none & $\leq 19.3$ & 225 & \phn30 & 2  & 13  \\
All giants & $\sigma_{M}$, $\sigma_{T_2}$, $\sigma_{DDO51} \leq 0.1$ &
$\leq 19.3$ & 223 & \phn29 & 2 & 13  \\

Extratidal giants & none & $\leq 20.8$ & 137 & \phn60 & 3 & 44  \\
Extratidal giants & $\sigma_{M}$, $\sigma_{T_2}$, $\sigma_{DDO51} \leq 0.1$ &
$\leq 20.8$ & 118 & \phn48 & 2 & 41  \\
Extratidal giants & none & $\leq 20.3$ & 111 & \phn48 & 2 & 43 \\
Extratidal giants & $\sigma_{M}$, $\sigma_{T_2}$, $\sigma_{DDO51} \leq 0.1$ &
$\leq 20.3$ & \phn90 & \phn35 & 2 & 39  \\
Extratidal giants & none & $\leq 19.8$ & \phn58 & \phn23 & 2 & 40  \\
Extratidal giants & $\sigma_{M}$, $\sigma_{T_2}$, $\sigma_{DDO51} \leq 0.1$ &
$\leq 19.8$ & \phn55 & \phn20 & 2 & 36  \\
Extratidal giants & none & $\leq 19.3$ & \phn29 & \phn11 & 1 & 38  \\
Extratidal giants & $\sigma_{M}$, $\sigma_{T_2}$, $\sigma_{DDO51} \leq 0.1$ &
$\leq 19.3$ & \phn27 & \phn10 & 1 & 37 \\

\enddata

\tablenotetext{a}{``All giants'' refers to a sample containing all giant
  candidates, whereas ``Extratidal giants'' refers to those samples that
  contain only candidate giants found outside the nominal tidal radius
  of Carina.}

\tablenotetext{b}{In \citeauthor{paperII}, we selected only those stars
  that had photometric errors $\leq 0.1$ in each of the three filters.
  The table compares the expected level of contamination between samples
  with no error cut (``none'') or with $\sigma_M$, $\sigma_{T_2}$,
  $\sigma_{DDO51} \leq 0.1$.  }

\tablenotetext{c}{$\langle N_{c} \rangle$ and $\sigma_{\langle N_c
    \rangle}$ have been rounded to the nearest star.}

\tablenotetext{d}{This is the error in the expected number of
  contaminants, $\langle N_{c} \rangle$, which is the sum of the
  numerical error in evaluating equation~(\ref{pint}) added in
  quadrature with the statistical error in the total probability (see \S
  3).}

\tablenotetext{e}{The fraction of $N_{\rm stars}$ expected to be
  contaminants.}

\label{contamprob}

\end{deluxetable}

\begin{deluxetable}{llccccc}
\tablewidth{480pt}
\tabletypesize{\small}

\tablecaption{Comparison of the photometric and spectroscopic Carina membership fraction}

\tablehead{ \colhead{Sample name} &
\colhead{Error Cut} &
\colhead{$M$ Limit}& 
\colhead{\%CM(\S2)\tablenotemark{a}} & 
\colhead{\%CM(PII)\tablenotemark{b}} &
\colhead{\%CM(spec)\tablenotemark{c}} &
\colhead{N(spec)}}   

\startdata
   
All giants & none & $\leq 20.8$ & $71\pm0.7$ & \nodata & 81 & 74 \\
All giants & $\sigma_{M}$, $\sigma_{T_2}$, $\sigma_{DDO51} \leq 0.1$ &
$\leq 20.8$ & $73\pm0.6$ & $96\pm1.3$ & 82 & 72 \\
All giants & none & $\leq 20.3$ & $76\pm0.7$ & \nodata & 85 & 66 \\
All giants & $\sigma_{M}$, $\sigma_{T_2}$, $\sigma_{DDO51} \leq 0.1$ &
$\leq 20.3$ & $79\pm0.7$ & $96\pm1.1$ & 85 & 66 \\
All giants & none & $\leq 19.8$ & $80\pm0.8$ & \nodata & 88 & 61 \\
All giants & $\sigma_{M}$, $\sigma_{T_2}$, $\sigma_{DDO51} \leq 0.1$ &
$\leq 19.8$ & $81\pm0.8$ & $96\pm0.8$ & 88 & 61 \\
All giants & none & $\leq 19.3$ & $87\pm0.9$ & \nodata & 94 & 53 \\
All giants & $\sigma_{M}$, $\sigma_{T_2}$, $\sigma_{DDO51} \leq 0.1$ &
$\leq 19.3$ & $87\pm0.9$ & $96\pm0.6$ & 94 & 53 \\

Extratidal giants & none & $\leq 20.8$ & $56\pm2.2$ & \nodata & 44 & 18 \\
Extratidal giants & $\sigma_{M}$, $\sigma_{T_2}$, $\sigma_{DDO51} \leq 0.1$ &
$\leq 20.8$ & $59\pm1.7$ & $73\pm11\phd$ & 44 & 18 \\
Extratidal giants & none & $\leq 20.3$ & $57\pm1.8$ & \nodata & 47 & 15 \\
Extratidal giants & $\sigma_{M}$, $\sigma_{T_2}$, $\sigma_{DDO51} \leq 0.1$ &
$\leq 20.3$ & $61\pm2.2$ & $76\pm6\phd\phn$ & 47 & 15 \\
Extratidal giants & none & $\leq 19.8$ & $60\pm3.4$ & \nodata & 58 & 12 \\
Extratidal giants & $\sigma_{M}$, $\sigma_{T_2}$, $\sigma_{DDO51} \leq 0.1$ &
$\leq 19.8$ & $64\pm3.6$ & $80\pm4\phd\phn$ & 58 & 12 \\
Extratidal giants & none & $\leq 19.3$ & $62\pm3.4$ & \nodata & 86 & \phn7 \\
Extratidal giants & $\sigma_{M}$, $\sigma_{T_2}$, $\sigma_{DDO51} \leq 0.1$ &
$\leq 19.3$ & $63\pm3.7$ & $78\pm4\phd\phn$ & 86 & \phn7 \\

\enddata

\label{contamcomp}   
\tablenotetext{a}{Membership Fraction predicted in \S2}
\tablenotetext{b}{Membership Fraction predicted in \citefullauthor{paperII}}
\tablenotetext{c}{Membership Fraction among spectroscopically observed candidates}
\end{deluxetable}

\begin{deluxetable}{lrcccr}
\tabletypesize{\small}
\tablewidth{0pt}
\tablecaption{Stars in Carina Field with Repeat Hydra Observations}
\tablehead{
\colhead{Star Name}& \colhead{$V_{\rm r}$} &  \colhead{$CCP$} & \colhead{$Q$} &
\colhead{$s(RV)$ }&\colhead{RV date} }
\startdata
    C88 &$\mathbf{  217.0}$ & \nodata& 7 & \phn3.9 & \nodata \\
\nodata &  213.8 &   0.47 & 7 & \nodata & 29Mar2000 \\
\nodata &  215.3 &   1.20 & 7 & \nodata & 08Oct2001 \\
\nodata &  221.1 &   0.62 & 7 & \nodata & 07Oct2001 \\
   C554 &$\mathbf{  229.5}$ & \nodata& 7 & \phn7.0 & \nodata \\
\nodata &  222.5 &   0.48 & 6 & \nodata & 07Oct2001 \\
\nodata &  230.3 &   1.24 & 7 & \nodata & 08Oct2001 \\
   C620 &$\mathbf{  252.4}$ & \nodata& 7 & \phn3.5 & \nodata \\
\nodata &  249.4 &   0.54 & 6 & \nodata & 29Mar2000 \\
\nodata &  254.2 &   0.35 & 4 & \nodata & 08Oct2001 \\
  C1394 &$\mathbf{  235.4}$ & \nodata& 5 &96.5 & \nodata \\
\nodata &  138.9 &   0.26 & 4 & \nodata & 09Oct2001 \\
\nodata &  238.7 &   0.55 & 5 & \nodata & 29Mar2000 \\
  C2362 &$\mathbf{  228.1}$ & \nodata& 7 & \phn0.7 & \nodata \\
\nodata &  227.6 &   0.54 & 6 & \nodata & 29Mar2000 \\
\nodata &  228.6 &   0.87 & 7 & \nodata & 30Mar2000 \\
  C3071 &$\mathbf{  229.0}$ & \nodata& 7 &28.9 & \nodata \\
\nodata &  201.1 &   0.25 & 4 & \nodata & 28Mar2000 \\
\nodata &  236.0 &   0.42 & 7 & \nodata & 30Mar2000 \\
C1215090&$\mathbf{   52.5}$ & \nodata& 7 & \phn3.6 & \nodata \\
\nodata &   51.6 &   0.59 & 6 & \nodata & 30Mar2000 \\
\nodata &   56.0 &   0.49 & 4 & \nodata & 28Mar2000 \\
C1404486&$\mathbf{  229.3}$ & \nodata& 7 & \phn6.6 & \nodata \\
\nodata &  223.0 &   0.47 & 6 & \nodata & 07Oct2001 \\
\nodata &  231.4 &   1.02 & 7 & \nodata & 08Oct2001 \\
C1406213&$\mathbf{  234.0}$ & \nodata& 7 & \phn4.1 & \nodata \\
\nodata &  230.3 &   0.36 & 6 & \nodata & 07Oct2001 \\
\nodata &  234.4 &   1.13 & 7 & \nodata & 08Oct2001 \\
C1406519&$\mathbf{  223.5}$ & \nodata& 7 & \phn1.1 & \nodata \\
\nodata &  222.2 &   0.65 & 7 & \nodata & 07Oct2001 \\
\nodata &  223.0 &   1.07 & 7 & \nodata & 30Mar2000 \\
\nodata &  224.2 &   1.14 & 7 & \nodata & 08Oct2001 \\
C1407251&$\mathbf{  230.2}$ & \nodata& 7 &10.5 & \nodata \\
\nodata &  228.5 &   0.70 & 7 & \nodata & 07Oct2001 \\
\nodata &  240.6 &   0.72 & 7 & \nodata & 29Mar2000 \\
C2201920&$\mathbf{   20.8}$ & \nodata& 7 &18.2 & \nodata \\
\nodata &    3.1 &   0.37 & 5 & \nodata & 30Mar2000 \\
\nodata &   25.2 &   0.55 & 7 & \nodata & 28Mar2000 \\
C2406923&$\mathbf{   45.2}$ & \nodata& 7 &13.0 & \nodata \\
\nodata &   32.6 &   0.32 & 5 & \nodata & 07Oct2001 \\
\nodata &   48.3 &   0.56 & 7 & \nodata & 08Oct2001 \\
C2408204&$\mathbf{  225.8}$ & \nodata& 7 & \phn4.1 & \nodata \\
\nodata &  224.8 &   0.87 & 7 & \nodata & 08Oct2001 \\
\nodata &  229.8 &   0.39 & 6 & \nodata & 07Oct2001 \\
C2411078&$\mathbf{  215.8}$ & \nodata& 7 & \phn1.9 & \nodata \\
\nodata &  214.0 &   0.23 & 4 & \nodata & 07Oct2001 \\
\nodata &  216.3 &   0.40 & 6 & \nodata & 08Oct2001 \\
C2415138&$\mathbf{  228.9}$ & \nodata& 7 &15.4 & \nodata \\
\nodata &  225.2 &   0.69 & 7 & \nodata & 08Oct2001 \\
\nodata &  243.9 &   0.46 & 7 & \nodata & 07Oct2001 \\
C2500670&$\mathbf{   -5.8}$ & \nodata& 7 & \phn5.0 & \nodata \\
\nodata & $-$2.3 &   0.84 & 7 & \nodata & 08Oct2001 \\
\nodata & $-$9.3 &   0.68 & 7 & \nodata & 07Oct2001 \\
C2501583&$\mathbf{  231.7}$ & \nodata& 7 &24.8 & \nodata \\
\nodata &  207.6 &   0.31 & 5 & \nodata & 07Oct2001 \\
\nodata &  237.7 &   0.98 & 7 & \nodata & 08Oct2001 \\
C2502058&$\mathbf{  225.0}$ & \nodata& 7 &20.0 & \nodata \\
\nodata &  205.0 &   0.30 & 4 & \nodata & 28Mar2000 \\
\nodata &  225.8 &   0.77 & 7 & \nodata & 08Oct2001 \\
C2502249&$\mathbf{  360.5}$ & \nodata& 7 & \phn0.4 & \nodata \\
\nodata &  360.2 &   0.54 & 6 & \nodata & 07Oct2001 \\
\nodata &  360.7 &   1.15 & 7 & \nodata & 08Oct2001 \\
C2502565&$\mathbf{   63.4}$ & \nodata& 7 & \phn7.6 & \nodata \\
\nodata &   63.1 &   0.61 & 7 & \nodata & 08Oct2001 \\
\nodata &   71.0 &   0.41 & 6 & \nodata & 28Mar2000 \\
C2503083&$\mathbf{  222.1}$ & \nodata& 7 & \phn3.0 & \nodata \\
\nodata &  220.6 &   0.32 & 4 & \nodata & 07Oct2001 \\
\nodata &  222.2 &   0.88 & 7 & \nodata & 08Oct2001 \\
\nodata &  226.0 &   0.26 & 4 & \nodata & 09Oct2001 \\
C2503385&$\mathbf{  220.0}$ & \nodata& 7 & \phn6.0 & \nodata \\
\nodata &  218.5 &   0.85 & 7 & \nodata & 08Oct2001 \\
\nodata &  225.8 &   0.32 & 5 & \nodata & 07Oct2001 \\
\enddata
\end{deluxetable}

\begin{deluxetable}{lccccc}
\tabletypesize{\small}
\tablewidth{0pt}
\tablecaption{Comparison between Hydra and Other Observations}
\tablehead{
\colhead{Star Name}& \colhead{$V_{\rm r}$} &  \colhead{$CCP$}  &
\colhead{$Q$ }&
\colhead{$\sigma_{\rm RV}$}& \colhead{Reference} }
\startdata
  C1547    & 218.2  & 0.51 & 7 & \nodata  & 1 \\
  Mateo16  & 221.3  & \nodata & \nodata &   3.0 & 2 \\
&&&&&\\
  C2282    & 225.2  & 0.66 & 7 & \nodata  & 1 \\
  Mateo12  & 221.9  & \nodata & \nodata &   2.7 & 2 \\
&&&&&\\
  C2774    & 206.7  & 0.31 & 5 & \nodata  & 1 \\
  Mateo11  & 223.7  & \nodata & \nodata &   3.0 & 2  \\
&&&&&\\
C2501583   & 237.7  & 0.98 & 7 & \nodata  & 1 \\
C2501583   & 287.4  & 0.24 & \nodata & \nodata  & 3 \\
&&&&&\\
C2501927   & 223.2  & 0.34 & 4 & \nodata & 1 \\
C2501927   & 223.1  & 0.20 & \nodata & \nodata  & 3 \\
&&&&&\\
C2103156   & 231.0  & 0.91 & 7 & \nodata  & 1 \\
C2103156   & 250.7  & 0.61 & \nodata & \nodata  & 3 \\ 
&&&&&\\
C1407251   & 228.5  & 0.70 & 7 & \nodata  & 1 \\
C1407251   & 233.1  & 0.77 & \nodata & \nodata  & 3 \\ 
\enddata
\tablerefs{(1) This Paper; (2)\citet{mat93}; (3) \citet{paperII}}
\end{deluxetable}

\begin{deluxetable}{lccccccccccc} 
\tabletypesize{\scriptsize}
\tabcolsep1.5mm
\tablewidth{0pt}
\tablecaption{Radial Velocities of Stars in the Carina Field }
\tablehead{ \colhead{Star} &
\colhead{$\alpha_{2000}$} &
\colhead{$\delta_{2000}$} &
\colhead{UT Date } &
\colhead{$M_o$ } &
\colhead{$\!\!\!(M-T_2)_o\!\!\!$ } &
\colhead{$\!\!\!(M-DDO51)_o\!\!\!$ } &
\colhead{$\!\!V_{\rm r}\!\!$ } &
\colhead{$CCP$} &
\colhead{Q}&
\colhead{Cand} &
\colhead{Mem?}}
\startdata
C3509707 & 6:36:51.12  & $-$51:12:57.6  & 07Oct2001         & 15.58 & 0.10 &$-$0.77\phs &  220.0 & 0.33 &  6 & N & ?\tablenotemark{b} \\
C2500036 & 6:37:12.00  & $-$51:03:18.0  & 28Mar2000         & 19.12 & 1.22 &  0.01 &  350.2 & 0.31 &  4 & Y & N \\
C2500670 & 6:37:40.44  & $-$51:05:02.4  & \tablenotemark{a} & 14.03 & 1.35 &$-$0.03\phs &  \phn\phn$-$5.8\phs & \nodata &  7 & N & N \\
C2501002 & 6:37:56.64  & $-$51:04:19.2  & 08Oct2001         & 16.73 & 1.15 &  0.02 &  234.1 & 0.58 &  6 & N & \phm{?}Y?\tablenotemark{b} \\
C2406923 & 6:38:04.92  & $-$50:57:57.6  & \tablenotemark{a} & 18.32 & 1.26 &  0.02 &   \phn45.2 & \nodata &  7 & N & N \\
C2501272 & 6:38:07.08  & $-$51:18:39.6  & 28Mar2000         & 17.04 & 1.07 &  0.04 &  153.8 & 0.51 &  6 & N & N \\
C2408204 & 6:38:18.60  & $-$50:55:30.0  & \tablenotemark{a} & 18.49 & 1.44 &  0.05 &  225.8 & \nodata &  7 & Y & Y \\
C2501583 & 6:38:22.56  & $-$51:10:58.8  & \tablenotemark{a} & 18.32 & 1.68 &  0.02 &  231.7 & \nodata &  7 & Y & Y \\
C2408784 & 6:38:24.00  & $-$50:56:27.6  & 08Oct2001         & 18.54 & 1.56 &  0.09 &  218.2 & 0.47 &  6 & N & Y\tablenotemark{c} \\
C2408672 & 6:38:25.08  & $-$50:41:45.6  & 09Oct2001         & 19.61 & 0.92 &$-$0.30\phs &  200.4 & 0.36 &  4 & N & N \\
C2501927 & 6:38:36.96  & $-$51:16:22.8  & 08Oct2001         & 18.50 & 1.58 &  0.04 &  223.2 & 0.34 &  4 & Y & Y \\
C2502058 & 6:38:40.92  & $-$51:23:09.6  & \tablenotemark{a} & 15.65 & 0.50 &$-$0.01\phs &  225.0 & \nodata &  7 & N & ?\tablenotemark{b} \\
C2502062 & 6:38:43.44  & $-$51:10:37.2  & 08Oct2001         & 19.35 & 1.27 &$-$0.02\phs &  169.4 & 0.29 &  4 & Y & N \\
C2410759 & 6:38:43.80  & $-$50:49:58.8  & 09Oct2001         & 20.51 & 1.41 &  0.05 &  234.9 & 0.83 &  7 & N & Y\tablenotemark{c} \\
C2411078 & 6:38:47.04  & $-$50:50:31.2  & \tablenotemark{a} & 18.65 & 1.42 &  0.04 &  215.8 & \nodata &  7 & Y & Y \\
C2502249 & 6:38:52.08  & $-$51:04:40.8  & \tablenotemark{a} & 17.29 & 1.20 &  0.05 &  360.5 & \nodata &  7 & N & N \\
C1401432 & 6:38:54.60  & $-$51:04:01.2  & 09Oct2001         & 20.17 & 1.25 &  0.06 &   \phn76.6 & 1.12 &  7 & Y & N \\
C2300060 & 6:38:58.20  & $-$50:26:27.6  & 30Mar2000         & 14.88 & 1.27 &$-$0.01\phs &  204.2 & 1.14 &  7 & N & N\tablenotemark{d} \\
C2502565 & 6:39:02.16  & $-$51:14:27.6  & \tablenotemark{a} & 14.82 & 0.22 &  0.01 &   \phn63.4 & \nodata &  7 & N & N \\
C2502589 & 6:39:03.24  & $-$51:13:26.4  & 08Oct2001         & 20.30 & 1.25 &  0.02 &  181.4 & 0.28 &  5 & Y & Y \\
C2413772 & 6:39:12.96  & $-$50:41:45.6  & 30Mar2000         & 18.50 & 1.40 &  0.01 &  215.6 & 0.45 &  6 & Y & Y \\
C1403017 & 6:39:14.76  & $-$51:03:43.2  & 09Oct2001         & 20.59 & 1.11 &  0.07 &  182.6 & 0.52 &  7 & Y & Y \\
C2503100 & 6:39:17.28  & $-$51:16:55.2  & 08Oct2001         & 17.94 & 1.09 &  0.04 &  118.8 & 0.51 &  7 & N & N \\
C2503083 & 6:39:18.72  & $-$51:06:32.4  & \tablenotemark{a} & 18.07 & 1.65 &  0.01 &  222.1 & \nodata &  7 & Y & Y \\
C2415138 & 6:39:23.76  & $-$50:52:33.6  & \tablenotemark{a} & 19.05 & 1.35 &  0.09 &  228.9 & \nodata &  7 & Y & Y \\
C1403884 & 6:39:26.64  & $-$50:58:15.6  & 09Oct2001         & 19.97 & 1.16 &  0.14 &  228.1 & 0.98 &  7 & Y & Y \\
C1403975 & 6:39:27.00  & $-$51:02:16.8  & 09Oct2001         & 20.43 & 1.07 &  0.13 &  241.0 & 0.95 &  7 & Y & Y \\
C2503385 & 6:39:29.88  & $-$51:09:18.0  & \tablenotemark{a} & 18.26 & 1.59 &  0.03 &  220.0 & \nodata &  7 & Y & Y \\
C1404486 & 6:39:35.28  & $-$50:51:50.4  & \tablenotemark{a} & 18.32 & 1.55 &  0.03 &  229.3 & \nodata &  7 & Y & Y \\
C2503632 & 6:39:37.44  & $-$51:13:08.4  & 08Oct2001         & 17.15 & 1.14 &  0.02 &  126.0 & 0.79 &  7 & N & N \\
C1404834 & 6:39:38.16  & $-$51:02:52.8  & 08Oct2001         & 18.74 & 1.24 &  0.03 &  236.4 & 0.42 &  5 & N & Y\tablenotemark{c} \\
C1405483 & 6:39:47.16  & $-$50:57:43.2  & 08Oct2001         & 18.53 & 1.38 &  0.06 &  223.0 & 0.22 &  4 & Y & Y \\
C1405730 & 6:39:51.84  & $-$50:47:45.6  & 30Mar2000         & 18.31 & 1.06 &  0.04 &  115.8 & 0.30 &  5 & N & N \\
C2301189 & 6:39:53.64  & $-$50:21:07.2  & 30Mar2000         & 20.53 & 1.07 &  0.10 &  752.6 & 0.22 &  4 & Y & N \\
C1406213 & 6:39:55.80  & $-$50:57:36.0  & \tablenotemark{a} & 18.05 & 1.60 &  0.02 &  234.0 & \nodata &  7 & Y & Y \\
C1406519 & 6:40:00.84  & $-$50:50:09.6  & \tablenotemark{a} & 17.56 & 1.93 &  0.00 &  223.5 & \nodata &  7 & N & Y\tablenotemark{c} \\
C1407251 & 6:40:08.76  & $-$50:57:10.8  & \tablenotemark{a} & 17.62 & 1.95 &$-$0.01\phs &  230.2 & \nodata &  7 & N & Y\tablenotemark{c} \\
C1407921 & 6:40:16.32  & $-$51:00:18.0  & 08Oct2001         & 18.26 & 1.50 &  0.02 &  241.7 & 0.71 &  7 & Y & Y \\
     C88 & 6:40:31.08  & $-$50:55:22.8  & \tablenotemark{a} & 17.97 & 1.62 &  0.03 &  217.0 & \nodata &  7 & Y & Y \\
C3009947 & 6:40:31.08  & $-$50:59:13.2  & 29Mar2000         & 17.99 & 1.63 &  0.01 &  206.8 & 0.40 &  6 & Y & Y \\
C2302163 & 6:40:37.20  & $-$50:37:37.2  & 12Nov2000         & 17.63 &$-$0.34\phs &$-$0.02\phs &  \phn\phn$-$4.4\phs & 0.70 &  7 & N & N \\
C2302255 & 6:40:40.80  & $-$50:38:13.2  & 30Mar2000         & 16.53 & 1.30 &$-$0.03\phs &  145.0 & 0.87 &  7 & N & N \\
C1609133 & 6:40:46.20  & $-$51:15:14.4  & 29Mar2000         & 14.52 & 0.50 &  0.02 &   \phn26.8 & 0.59 &  7 & N & N \\
    C554 & 6:40:46.56  & $-$51:01:40.8  & \tablenotemark{a} & 17.87 & 1.63 &  0.02 &  229.5 & \nodata &  7 & Y & Y \\
    C620 & 6:40:47.64  & $-$51:06:03.6  & 29Mar2000         & 17.77 & 1.69 &  0.00 &  252.4 & 0.00 &  7 & Y & Y \\
C1411419 & 6:40:54.84  & $-$50:44:02.4  & 30Mar2000         & 19.02 & 1.31 &  0.00 &  227.7 & 0.20 &  4 & Y & Y \\
    C918 & 6:40:58.08  & $-$51:02:27.6  & 29Mar2000         & 17.87 & 1.50 &  0.03 &  249.2 & 0.45 &  5 & N & Y\tablenotemark{c} \\
    C921 & 6:40:58.08  & $-$51:01:58.8  & 29Mar2000         & 18.09 & 1.52 &  0.04 &  242.9 & 0.49 &  5 & Y & Y \\
C2302805 & 6:41:02.76  & $-$50:19:58.8  & 30Mar2000         & 14.19 & 1.32 &$-$0.04\phs &   \phn48.4 & 1.14 &  7 & N & N \\
C2616011 & 6:41:03.12  & $-$51:27:57.6  & 29Mar2000         & 18.91 & 1.09 &  0.03 &  117.7 & 0.24 &  4 & N & N \\
   C1056 & 6:41:03.48  & $-$50:57:03.6  & 12Nov2000         & 18.19 & 1.50 &  0.01 &  232.6 & 0.31 &  4 & Y & Y \\
C1200281 & 6:41:03.84  & $-$50:46:04.8  & 09Oct2001         & 19.35 & 1.10 &  0.04 &  227.4 & 0.59 &  7 & N & Y\tablenotemark{c} \\
C3007069 & 6:41:04.92  & $-$51:01:33.6  & 29Mar2000         & 17.79 & 1.69 &  0.08 &  213.0 & 0.61 &  7 & Y & Y \\
   C1176 & 6:41:05.28  & $-$51:05:24.0  & 29Mar2000         & 17.78 & 1.69 &  0.02 &  241.0 & 0.59 &  5 & Y & Y \\
C2302966 & 6:41:06.36  & $-$50:38:09.6  & 30Mar2000         & 14.51 & 1.37 &$-$0.04\phs &   \phn47.5 & 0.89 &  7 & N & N \\
   C1201 & 6:41:07.44  & $-$50:59:24.0  & 12Nov2000         & 18.38 &$-$0.85\phs &$-$1.01\phs &  229.5 & 0.43 &  6 & N & Y \\
C2302947 & 6:41:07.80  & $-$50:25:04.8  & 30Mar2000         & 15.08 & 0.56 &$-$0.00\phs &   \phn24.7 & 0.88 &  7 & N & N \\
C1200940 & 6:41:09.60  & $-$50:40:08.4  & 12Nov2000         & 19.64 & 1.05 &  0.05 &  225.6 & 0.29 &  4 & N & Y\tablenotemark{c} \\
   C1291 & 6:41:10.68  & $-$50:55:51.6  & \tablenotemark{e} & 18.20 & 1.58 &  0.01 &  234.5 & \nodata &  M & Y & Y \\
C1613192 & 6:41:12.12  & $-$51:13:15.6  & 29Mar2000         & 18.15 & 1.56 &  0.06 &  249.2 & 0.41 &  5 & Y & Y \\
   C1394 & 6:41:14.64  & $-$50:51:10.8  & \tablenotemark{a} & 17.84 & 1.56 &  0.00 &  235.4 & \nodata &  5 & N & Y\tablenotemark{c} \\
   C1501 & 6:41:15.36  & $-$51:01:15.6  & 12Nov2000         & 18.37 & 1.50 &$-$0.03\phs &  221.1 & 0.29 &  4 & Y & Y \\
   C1495 & 6:41:15.72  & $-$50:59:49.2  & \tablenotemark{e} & 18.03 & 1.65 &  0.03 &  228.1 & \nodata &  M & Y & Y \\
C3001272 & 6:41:16.44  & $-$51:08:45.6  & 29Mar2000         & 20.84 & 0.12 &  0.12 &  176.8 & 0.19 &  4 & N & Y \\
   C1547 & 6:41:16.80  & $-$51:00:54.0  & 29Mar2000         & 17.98 & 1.64 &  0.05 &  218.2 & 0.51 &  7 & Y & Y \\
   C1552 & 6:41:18.60  & $-$50:53:13.2  & \tablenotemark{e} & 18.24 & 1.56 &  0.02 &  222.4 & \nodata &  M & Y & Y \\
   C1644 & 6:41:19.68  & $-$50:57:25.2  & \tablenotemark{e} & 18.20 & 1.55 &  0.02 &  222.5 & \nodata &  M & Y & Y \\
   C1685 & 6:41:20.40  & $-$51:02:06.0  & \tablenotemark{e} & 17.81 & 1.73 &$-$0.27\phs &  \phn\phn$-$1.3\phs & \nodata &  M & N & N \\
C2303229 & 6:41:21.48  & $-$50:17:02.4  & 30Mar2000         & 13.83 & 1.39 &$-$0.01\phs &    \phn\phn7.2 & 0.90 &  7 & N & N \\
   C1759 & 6:41:21.84  & $-$51:03:43.2  & 29Mar2000         & 17.97 & 1.59 &  0.00 &  227.4 & 0.48 &  5 & Y & Y \\
   C1792 & 6:41:23.28  & $-$51:00:50.4  & \tablenotemark{e} & 17.77 & 1.91 &$-$0.28\phs &   \phn83.9 & \nodata &  M & N & N \\
   C1862 & 6:41:26.16  & $-$50:55:44.4  & \tablenotemark{e} & 17.84 & 1.68 &  0.00 &  211.4 & \nodata &  M & Y & Y \\
   C1956 & 6:41:27.24  & $-$51:00:18.0  & \tablenotemark{e} & 17.93 & 1.66 &  0.03 &  236.3 & \nodata &  M & Y & Y \\
   C2186 & 6:41:33.72  & $-$50:55:33.6  & \tablenotemark{e} & 18.21 & 1.98 &$-$0.30\phs & \phn$-$11.4\phs & \nodata &  M & N & N \\
C1204278 & 6:41:34.08  & $-$50:32:20.4  & 30Mar2000         & 19.22 & 1.26 &$-$0.04\phs &  175.9 & 0.27 &  4 & N & N \\
   C2282 & 6:41:36.60  & $-$50:56:24.0  & 09Oct2001         & 18.12 & 1.61 &  0.03 &  225.2 & 0.66 &  7 & Y & Y \\
   C2252 & 6:41:36.96  & $-$50:50:06.0  & 30Mar2000         & 18.07 & 1.55 &  0.05 &  246.8 & 0.53 &  7 & Y & Y \\
   C2369 & 6:41:37.32  & $-$51:00:39.6  & \tablenotemark{e} & 18.23 & 1.60 &  0.02 &  214.7 & \nodata &  M & Y & Y \\
   C2396 & 6:41:37.68  & $-$51:01:44.4  & \tablenotemark{e} & 18.18 & 1.61 &  0.04 &  224.8 & \nodata &  M & Y & Y \\
   C2359 & 6:41:38.04  & $-$50:56:45.6  & \tablenotemark{e} & 18.00 & 1.64 &$-$0.35\phs &    \phn\phn3.8 & \nodata &  M & N & N \\
   C2362 & 6:41:39.48  & $-$50:49:58.8  & \tablenotemark{a} & 17.84 & 1.68 &  0.03 &  228.1 & \nodata &  7 & Y & Y \\
C1618075 & 6:41:42.36  & $-$51:13:55.2  & 29Mar2000         & 18.12 & 1.25 &$-$0.02\phs &   \phn71.8 & 0.46 &  6 & N & N \\
   C2563 & 6:41:44.16  & $-$50:50:16.8  & 09Oct2001         & 18.75 & 0.35 &$-$0.07\phs &  238.6 & 0.50 &  6 & N & Y \\
   C2719 & 6:41:46.32  & $-$50:58:55.2  & \tablenotemark{e} & 17.89 & 1.44 &$-$0.26\phs &   \phn23.7 & \nodata &  M & N & N \\
C3007367 & 6:41:46.32  & $-$51:01:22.8  & \tablenotemark{e} & 18.07 & 1.61 &$-$0.05\phs &  211.9 & \nodata &  M & Y & Y \\
   C2774 & 6:41:47.76  & $-$50:59:45.6  & 09Oct2001         & 18.29 & 1.50 &  0.01 &  206.2 & 0.31    &  5 & Y & Y \\
   C2764 & 6:41:48.12  & $-$50:55:01.2  & \tablenotemark{e} & 17.85 & 1.68 &  0.00 &  224.1 & \nodata &  M & Y & Y \\
C1207193 & 6:41:51.00  & $-$50:45:32.4  & 12Nov2000         & 18.78 & 1.46 &$-$0.02\phs &  264.7 & 0.38 &  5 & Y & N \\
C2200446 & 6:41:53.88  & $-$50:31:15.6  & 12Nov2000         & 20.73 & 0.88 &  0.33 &  232.6 & 0.31 &  4 & Y & Y \\
   C2995 & 6:41:54.60  & $-$50:57:00.0  & \tablenotemark{e} & 17.84 & 1.77 &$-$0.02\phs &  229.7 & \nodata &  M & Y & Y \\
   C3110 & 6:41:57.84  & $-$50:57:14.4  & \tablenotemark{e} & 18.08 & 1.52 &  0.01 &  228.7 & \nodata &  M & Y & Y \\
   C3132 & 6:41:57.84  & $-$50:59:52.8  & \tablenotemark{e} & 17.92 & 1.73 &  0.01 &  222.5 & \nodata &  M & Y & Y \\
   C3071 & 6:41:58.20  & $-$50:48:57.6  & \tablenotemark{a} & 18.63 & 1.30 &$-$0.01\phs &  229.0 & \nodata &  7 & Y & Y \\
C1208301 & 6:41:58.20  & $-$50:46:40.8  & 30Mar2000         & 18.73 & 1.47 &  0.03 &  211.2 & 0.20 &  4 & Y & Y \\
C1208389 & 6:41:58.20  & $-$50:49:48.0  & 12Nov2000         & 19.33 & 1.31 &  0.04 &  195.9 & 0.77 &  7 & Y & Y \\
C2200735 & 6:41:58.56  & $-$50:36:00.0  & 28Mar2000         & 15.89 & 0.58 &  0.02 &    \phn\phn2.0 & 0.66 &  7 & N & N \\
   C3135 & 6:41:59.64  & $-$50:51:14.4  & 30Mar2000         & 17.96 & 1.65 &  0.03 &  214.1 & 0.42 &  5 & Y & Y \\
   C3179 & 6:41:59.64  & $-$50:58:40.8  & \tablenotemark{e} & 17.98 & 1.49 &$-$0.28\phs &   \phn23.8 & \nodata &  M & N & N \\
   C3218 & 6:42:00.00  & $-$51:01:51.6  & \tablenotemark{e} & 17.86 & 1.75 &  0.02 &  210.0 & \nodata &  M & Y & Y \\
   C3277 & 6:42:01.08  & $-$51:03:43.2  & 29Mar2000         & 17.97 & 1.56 &  0.02 &  236.3 & 0.36 &  4 & Y & Y \\
C1623049 & 6:42:11.16  & $-$51:20:38.4  & 29Mar2000         & 19.70 & 0.95 &  0.18 &   \phn13.7 & 0.15 &  4 & N & N \\
   C3800 & 6:42:18.72  & $-$50:48:00.0  & 12Nov2000         & 19.63 & 1.32 &  0.07 &  184.2 & 0.22 &  4 & Y & Y \\
C1211401 & 6:42:19.44  & $-$50:42:39.6  & 12Nov2000         & 16.55 & 0.28 &  0.04 &  218.0 & 0.49 &  7 & N & \phm{?}Y?\tablenotemark{b} \\
   C3897 & 6:42:20.16  & $-$50:53:34.8  & 12Nov2000         & 16.86 & 0.70 &  0.03 &  229.5 & 0.43 &  5 & N & \phm{?}Y?\tablenotemark{b} \\
C2201920 & 6:42:20.88  & $-$50:32:34.8  & \tablenotemark{a} & 15.86 & 0.41 &$-$0.02\phs &   \phn20.8 & \nodata &  7 & N & N \\
C2201879 & 6:42:21.60  & $-$50:24:36.0  & 12Nov2000         & 19.42 & 1.28 &$-$0.03\phs &  220.9 & 0.28 &  4 & Y & Y \\
   C3994 & 6:42:23.04  & $-$50:52:26.4  & 28Mar2000         & 18.44 & 0.27 &  0.01 &  163.7 & 0.22 &  4 & N & Y \\
   C4156 & 6:42:25.20  & $-$51:03:46.8  & 12Nov2000         & 18.62 & 0.30 &$-$0.05\phs &  207.6 & 0.62 &  7 & N & Y \\
   C4228 & 6:42:28.44  & $-$51:00:03.6  & 12Nov2000         & 17.95 & 1.68 &  0.00 &  223.4 & 0.48 &  6 & Y & Y \\
   C4297 & 6:42:28.80  & $-$51:04:58.8  & 29Mar2000         & 18.92 & 1.35 &  0.07 &  187.5 & 0.25 &  4 & Y & Y \\
C1214844 & 6:42:41.40  & $-$50:47:42.0  & 12Nov2000         & 19.31 & 1.39 &  0.02 &  221.9 & 0.58 &  7 & Y & Y \\
C1214761 & 6:42:42.12  & $-$50:41:09.6  & 12Nov2000         & 19.44 & 1.26 &  0.08 &  213.4 & 0.37 &  4 & Y & Y \\
C1215090 & 6:42:44.28  & $-$50:40:26.4  & \tablenotemark{a} & 16.23 & 0.55 &  0.03 &   \phn52.5 & \nodata &  7 & N & N \\
C1806792 & 6:43:11.64  & $-$50:55:04.8  & 12Nov2000         & 18.61 & 1.47 &  0.05 &$-$125.4\phs & 0.32 &  5 & Y & N \\
C2206604 & 6:43:38.28  & $-$50:31:01.2  & 28Mar2000         & 16.21 & 1.33 &$-$0.03\phs &   \phn29.4 & 0.77 &  7 & N & N \\
C2100138 & 6:43:55.20  & $-$50:41:20.4  & 12Nov2000         & 20.17 & 1.34 &  0.07 &  232.0 & 0.35 &  4 & Y & Y \\
C2800436 & 6:43:57.00  & $-$51:07:44.4  & 13Nov2000         & 19.57 & 1.22 &  0.03 &   \phn22.9 & 0.47 &  7 & Y & N \\
C2802144 & 6:44:08.88  & $-$51:07:58.8  & 13Nov2000         & 17.82 & 1.10 &  0.02 &   \phn13.2 & 1.09 &  7 & N & N \\
C2100515 & 6:44:09.24  & $-$50:31:44.4  & 12Nov2000         & 18.59 & 1.43 &  0.04 &  194.6 & 1.00 &  7 & Y & Y \\
C2803049 & 6:44:12.48  & $-$51:21:25.2  & 13Nov2000         & 18.49 & 1.00 &  0.05 &  383.3 & 1.06 &  6 & N & N \\
C2804497 & 6:44:24.72  & $-$51:11:31.2  & 13Nov2000         & 20.18 & 1.00 &  0.19 & \phn$-$25.6\phs & 0.45 &  5 & Y & N \\
C2101462 & 6:44:35.88  & $-$50:47:56.4  & 28Mar2000         & 14.19 & 1.17 &  0.01 &   \phn79.1 & 0.90 &  7 & N & N \\ 
C2807582 & 6:44:45.24  & $-$51:15:10.8  & 13Nov2000         & 19.31 & 1.16 &  0.08 &  \phn\phn$-$8.2\phs & 0.67 &  7 & Y & N \\
C2807500 & 6:44:45.96  & $-$51:06:39.6  & 13Nov2000         & 18.19 & 0.44 &$-$0.02\phs &  \phn\phn$-$4.2\phs & 0.51 &  5 & N & N \\ 
C2101857 & 6:44:52.08  & $-$50:32:31.2  & 28Mar2000         & 15.45 & 1.24 &$-$0.01\phs &  127.3 & 0.99 &  7 & N & N \\
C2811401 & 6:45:11.88  & $-$51:16:58.8  & 13Nov2000         & 19.14 & 0.75 &  0.18 & \phn$-$49.6\phs & 0.63 &  7 & N & N \\ 
C2812177 & 6:45:18.36  & $-$51:12:00.0  & 13Nov2000         & 17.70 & 1.06 &  0.03 &   \phn15.0 & 0.24 &  4 & N & N \\
C2103156 & 6:45:30.96  & $-$50:49:26.4  & 29Mar2000         & 17.76 & 2.00 &$-$0.03\phs &  231.0 & 0.91 &  7 & N & Y\tablenotemark{c} \\
C4800378 & 6:45:59.40  & $-$51:27:14.4  & 13Nov2000         & 19.12 & 1.24 &  0.11 &   \phn28.6 & 0.41 &  4 & Y & N \\ 
C4800708 & 6:46:22.80  & $-$51:26:49.2  & 13Nov2000         & 19.32 & 1.26 &  0.04 & \phn$-$10.3\phs & 1.01 &  7 & Y & N \\ 
C4800961 & 6:46:41.52  & $-$51:28:40.8  & 13Nov2000         & 19.63 & 1.21 &  0.18 &    \phn\phn4.0 & 1.08 &  7 & Y & N \\ 
C4801898 & 6:47:34.44  & $-$51:08:38.4  & 13Nov2000         & 19.19 & 1.08 &  0.06 &$-$204.3\phs & 0.35 &  4 & N & N \\
C4801933 & 6:47:36.24  & $-$51:11:20.4  & 13Nov2000         & 19.27 & 1.05 &  0.06 &   \phn36.5 & 0.40 &  5 & N & N \\
C4801949 & 6:47:36.96  & $-$51:13:26.4  & 13Nov2000         & 19.69 & 1.36 &  0.00 & \phn$-$56.7\phs & 0.71 &  7 & Y & N \\ 
\enddata 

\tablenotetext{a}{Multiple Exposures, see Table 3.}

\tablenotetext{b}{Possible PAGB Carina star (see \S3.3.3.)} 

\tablenotetext{c}{Star was just outside our RGB selection boundary, but
  has the correct RV and is also selected as a giant in the 2CD.}

\tablenotetext{d}{Star has correct RV but is not considered a member
  because of location in the CMD.}

\tablenotetext{e}{Stars from \citet{mat93} not observed by us.}
\end{deluxetable}

\begin{figure}
\epsscale{0.8} 
\plotone{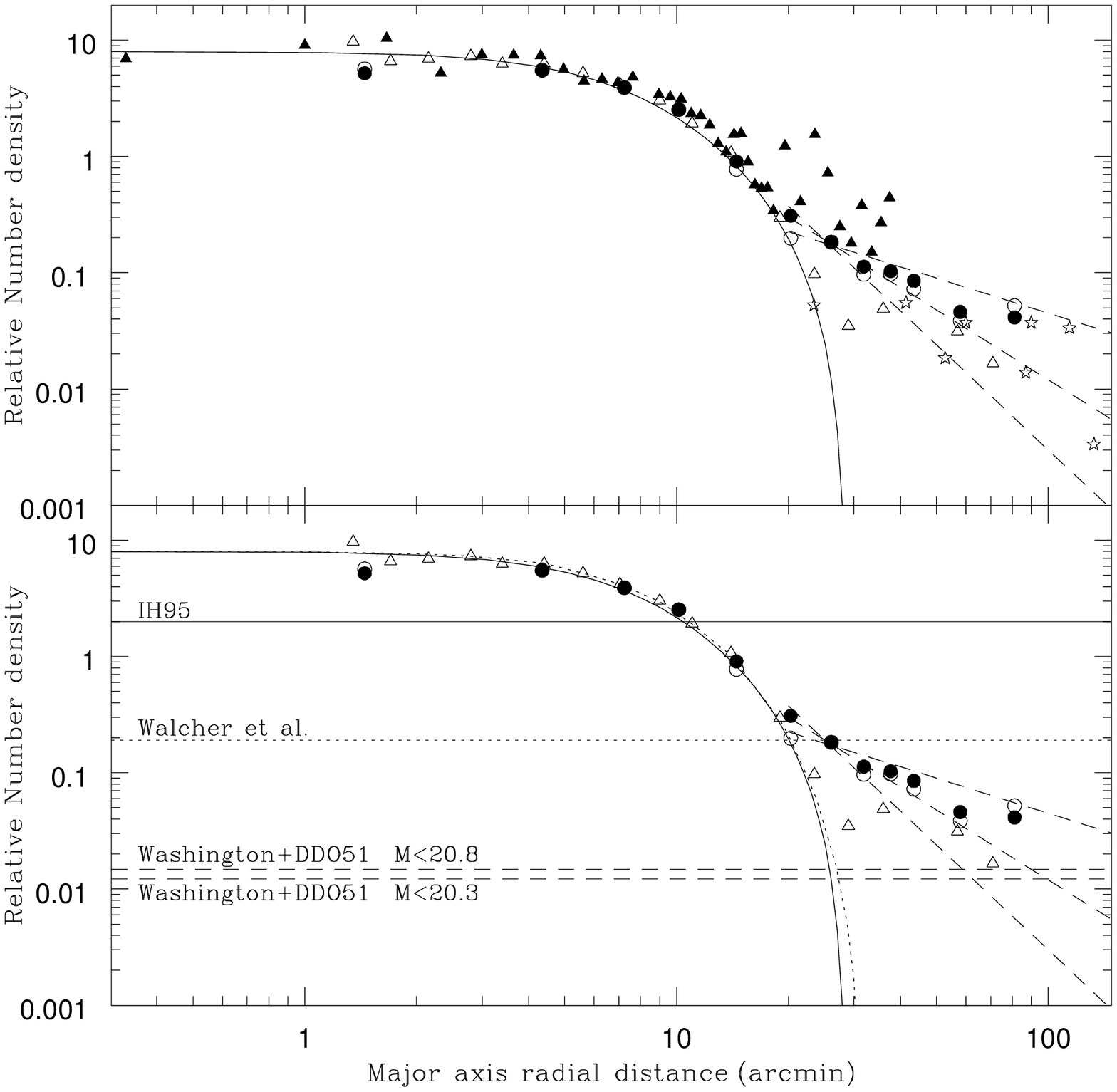}
\caption{Background-subtracted radial profile of the Carina dSph in a figure
  adapted from \citeauthor{paperII}.  In both panels the filled circles
  are from the \citeauthor{paperII} analysis to $M<20.8$ and the open
  circles are from the \citeauthor{paperII} analysis for $M<20.3$.  In
  the top panel we also show the points from \citeauthor{IH95} as solid
  triangles, the data points from \citet{KSH96} as stars, and the points
  from \citeauthor{W03} as open triangles (we do not show those points
  from \citeauthor{W03} that correspond to negative densities and which
  they have plotted in their corresponding plot for comparison).  In
  both panels, the solid curve is the King function fit from
  \citeauthor{IH95}, whereas the \citeauthor{W03} theoretical King
  function fit is shown as the dotted curve in the bottom panel.  The
  angled dashed lines in both panels are various power law functions in
  the extratidal region (see \citeauthor{paperII}).  In the bottom panel
  we also show the background levels estimated and subtracted from the
  measured densities by each survey to give the radial profiles.  The
  adopted \citeauthor{IH95} and \citeauthor{paperII} backgrounds are
  explicitly given in those publications; the \citeauthor{W03}
  background level was estimated based on the statement of these authors
  that the background was equal to the Carina profile at $20\arcmin$.
  Because of the varying survey magnitude limits, number densities have
  been normalized at $r=8\farcm3$ and the background levels scaled
  accordingly.  }
\end{figure}

\begin{figure}
\epsscale{1.0}
\plotone{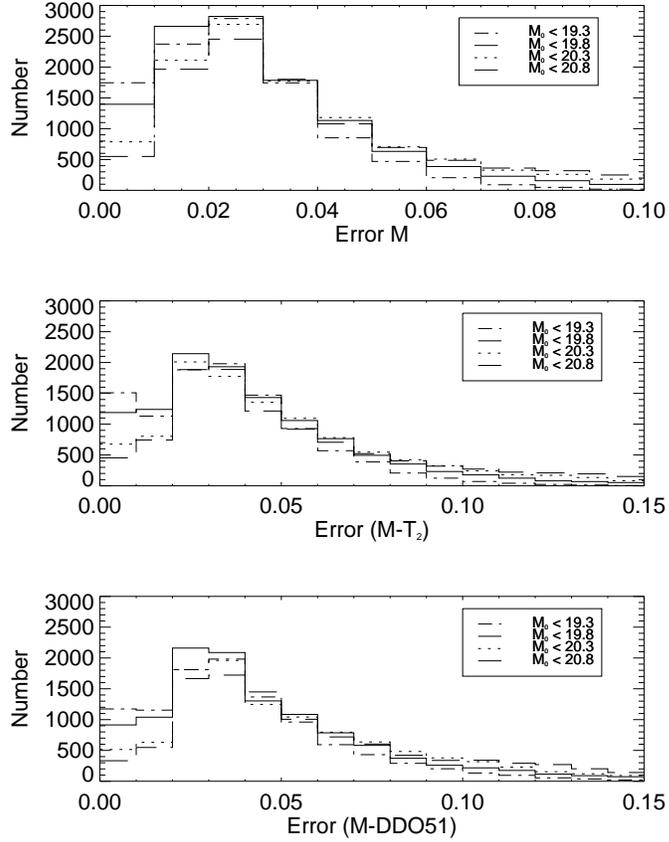}
\caption{Distribution of errors in the 0.10
  magnitude error-limited sample used in \citeauthor{paperII}.  For each
  panel, the distributions for each of the four magnitude-limited
  samples used in that analysis is shown.  It is important to note that
  the four magnitude-limited samples used in \citeauthor{paperII} each
  accounted for the loss of survey fields that were photometrically
  incomplete at each magnitude limit.  The fact that the error
  distributions are similarly shaped at each magnitude limit is a result
  of the fact that the survey fields lost at each magnitude limit are
  those with the worst magnitude errors }
\end{figure}

\begin{figure}
\epsscale{0.8}
\plotone{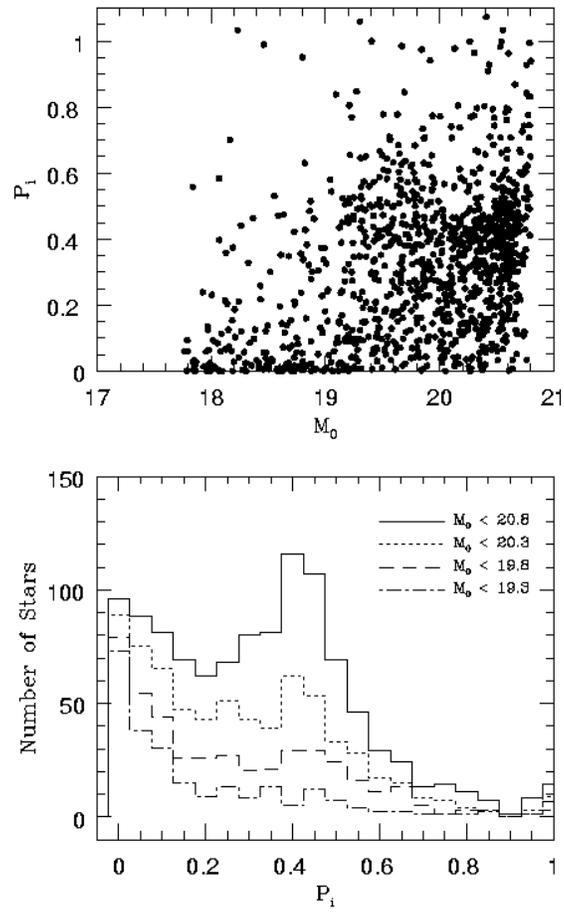}
\caption{(Top) Distribution of the probability of
  being a contaminant, $P_i$, as a function of $M$ magnitude. (Bottom)
  Distribution of $P_i$ for the four magnitude limits adopted in
  \citeauthor{paperII}.}
\end{figure}

\begin{figure}
\epsscale{1.0} 
\plotone{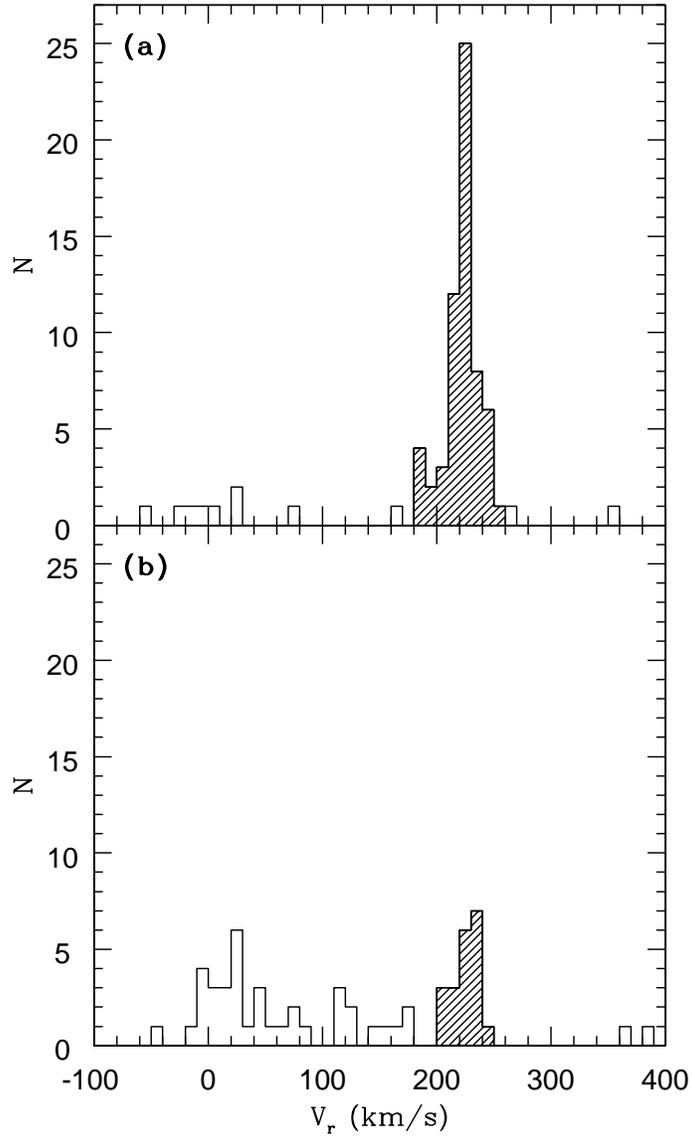}
\caption{Histograms of the radial velocities for all stars divided into
  (a) stars selected photometrically as Carina giant candidates, and (b)
  stars not photometrically selected as Carina giant candidates.  In the
  top panel the shaded region demarcates those stars considered to be
  Carina RV members.  In the bottom panel, the shaded points demarcate
  the {\it initial sample} from which we search for additional Carina
  members (see \S3.3).  }
\end{figure}

\begin{figure}
\epsscale{1.00} 
\plotone{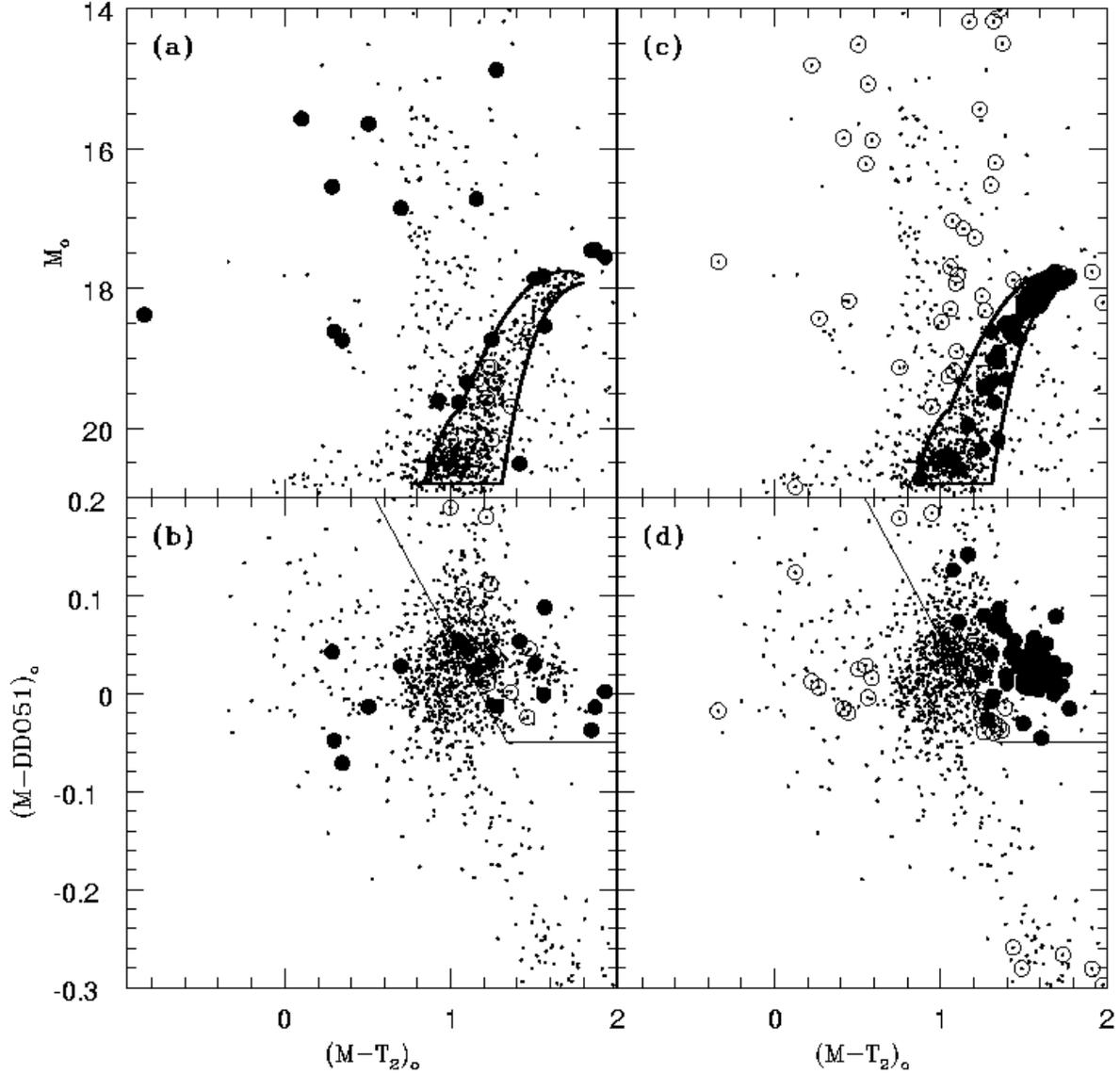} 
\epsscale{0.65}
\caption{(a) $(M - T_2,M)_o$ diagram marking the location of stars not
  photometrically selected as giants but with velocities consistent with
  being a Carina member (solid circles) and stars selected as giants but
  with a velocity not consistent with being a Carina member (open
  circles).  (b) $(M - T_2, M - DDO51,)_o$ diagram for the same data as
  panel (a). (c) $(M - T_2,M)_o$ diagram showing the location of stars
  selected as giant candidates and confirmed spectroscopically as Carina
  members (solid circles) as well as stars not selected to be Carina
  giants having RVs inconsistent with Carina membership (open circles).
  (d) $(M - T_2, M - DDO51,)_o$ diagram for the same data shown in panel
  (c).  In all panels, the dots show stars within 0.2 King limiting
  radii (\citeauthor{IH95}) and were plotted as a guide to the general
  CMD features of the Carina field.  The solid lines in all panels
  delineate the CMD and 2CD selection criteria.  The dwarf stars
  appearing in the lower right of panel (d) are from the \citet{mat93}
  study, and were not selected as giant stars in \citeauthor{paperII}.}
\epsscale{1.0}
\end{figure}

\begin{figure}
\epsscale{1.00} 
\plotone{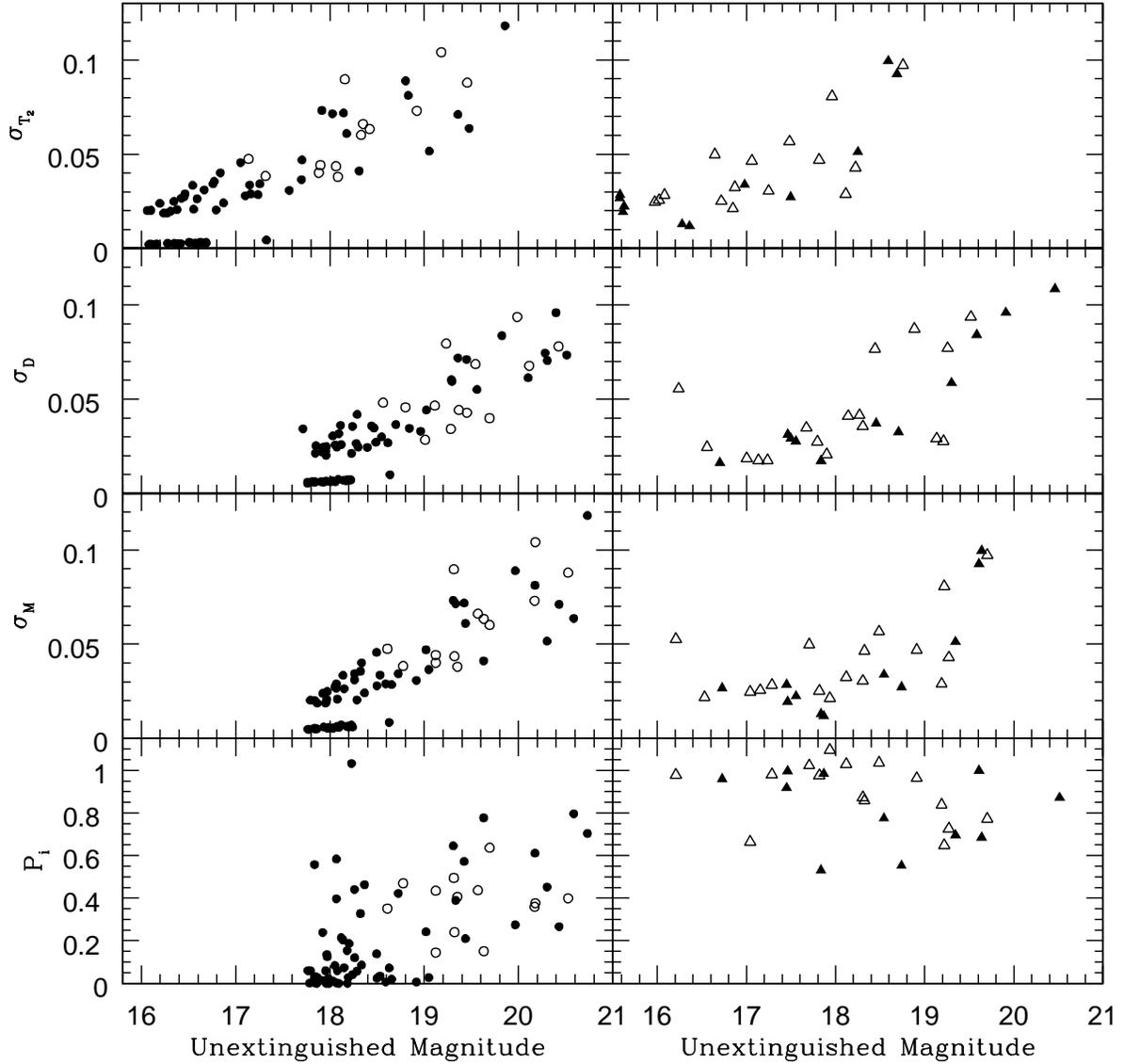} 
\caption{Distribution of photometric errors and contamination probabilities
  for those \citeauthor{paperII} stars having RVs: (left panels)
  selected photometrically as Carina giant candidates, and (right
  panels) those stars photometrically selected to be non-Carina giant
  candidates including stars lying just outside of the Carina giant
  selection criteria.  In all panels, stars with Carina-like RVs are
  shown with solid symbols and stars found to be RV non-members are
  shown with open symbols.}
\end{figure}

\begin{figure}
\epsscale{1.00}
\plotone{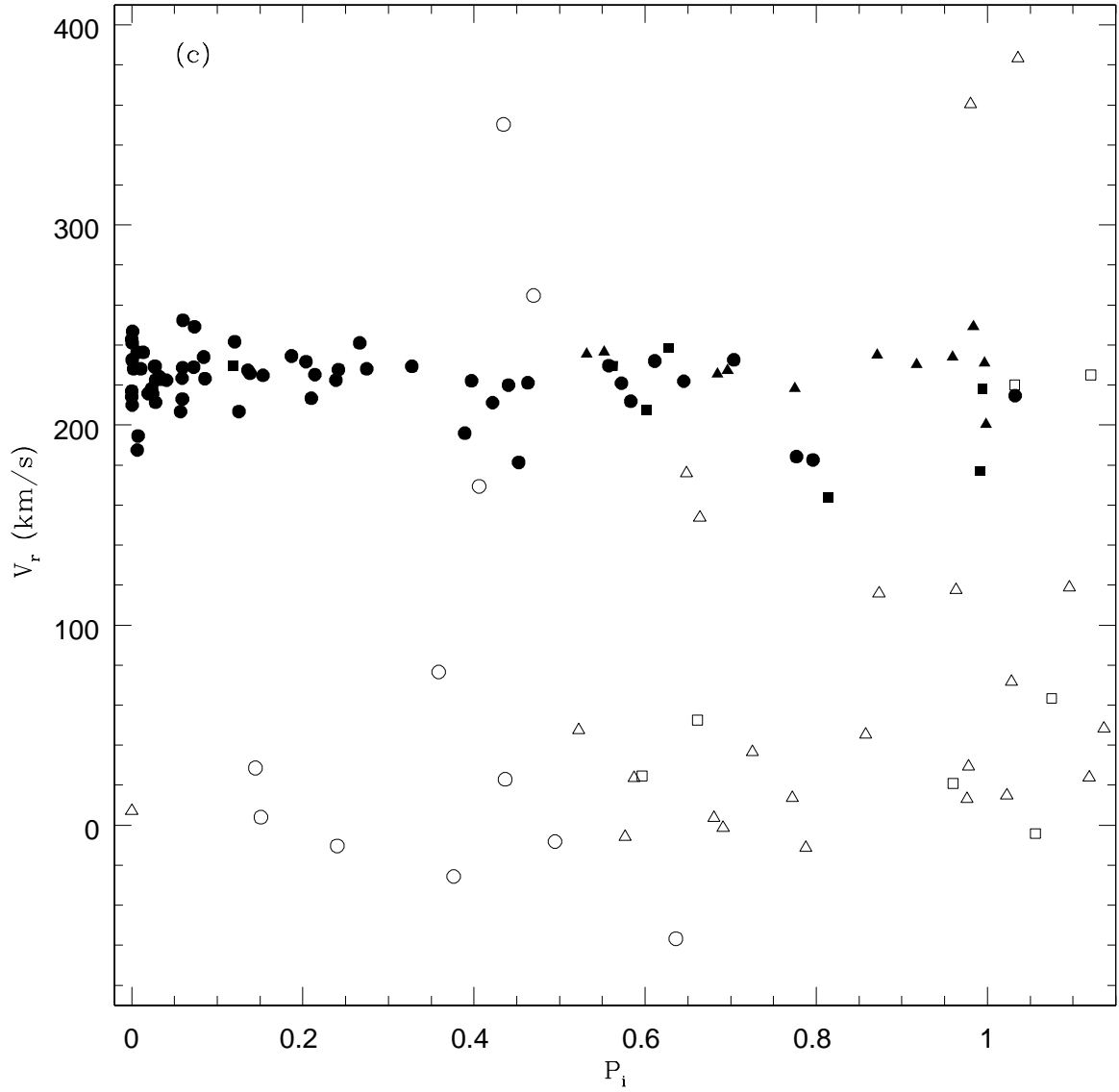}
\caption{Carina field RVs as a function of their $P_i$ probability for being a
  contaminant.  Circles show stars selected as Carina giants, squares
  are blue stars and all other stars are shown as triangles.  Filled
  symbols are used for objects considered to be Carina RV members.  }
\end{figure}

\begin{figure}
\epsscale{1.00}
\plotone{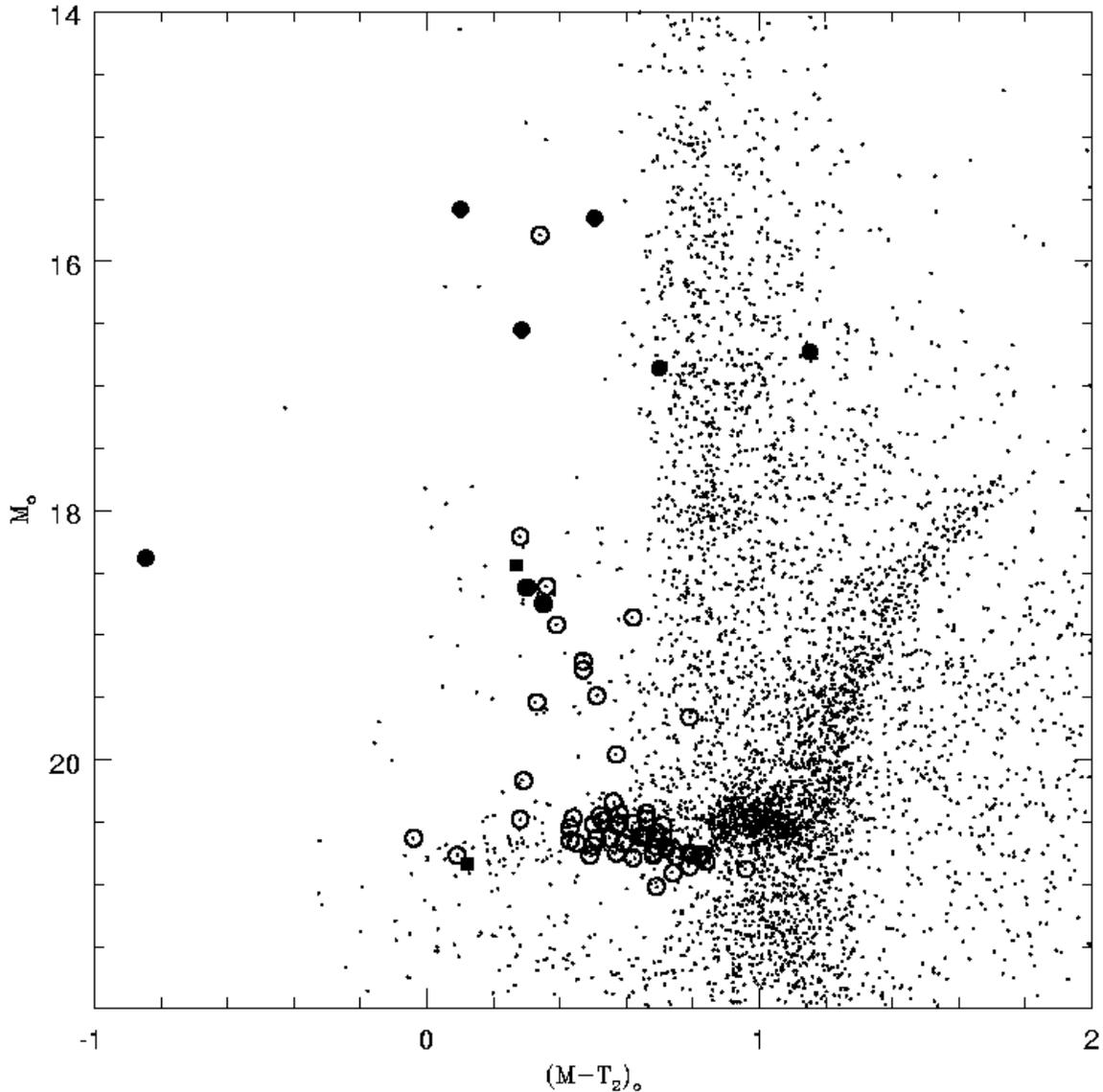}
\caption{Color-magnitude diagram to highlight spectroscopic results for 
  blue stars observed.  The small points are all stars in the
  photometric database, but limited to stars within 0.7 core radii for
  clarity.  Open circles mark stars found to be variables by
  \citet{ora03}, including both RR Lyraes and anomalous Cepheids.
  Filled circles are blue stars observed spectroscopically and found to
  have Carina-like RVs. Filled squares are blue stars with velocities
  slightly outside our membership criteria.}
\end{figure}

\begin{figure}
\epsscale{1.00}
\plotone{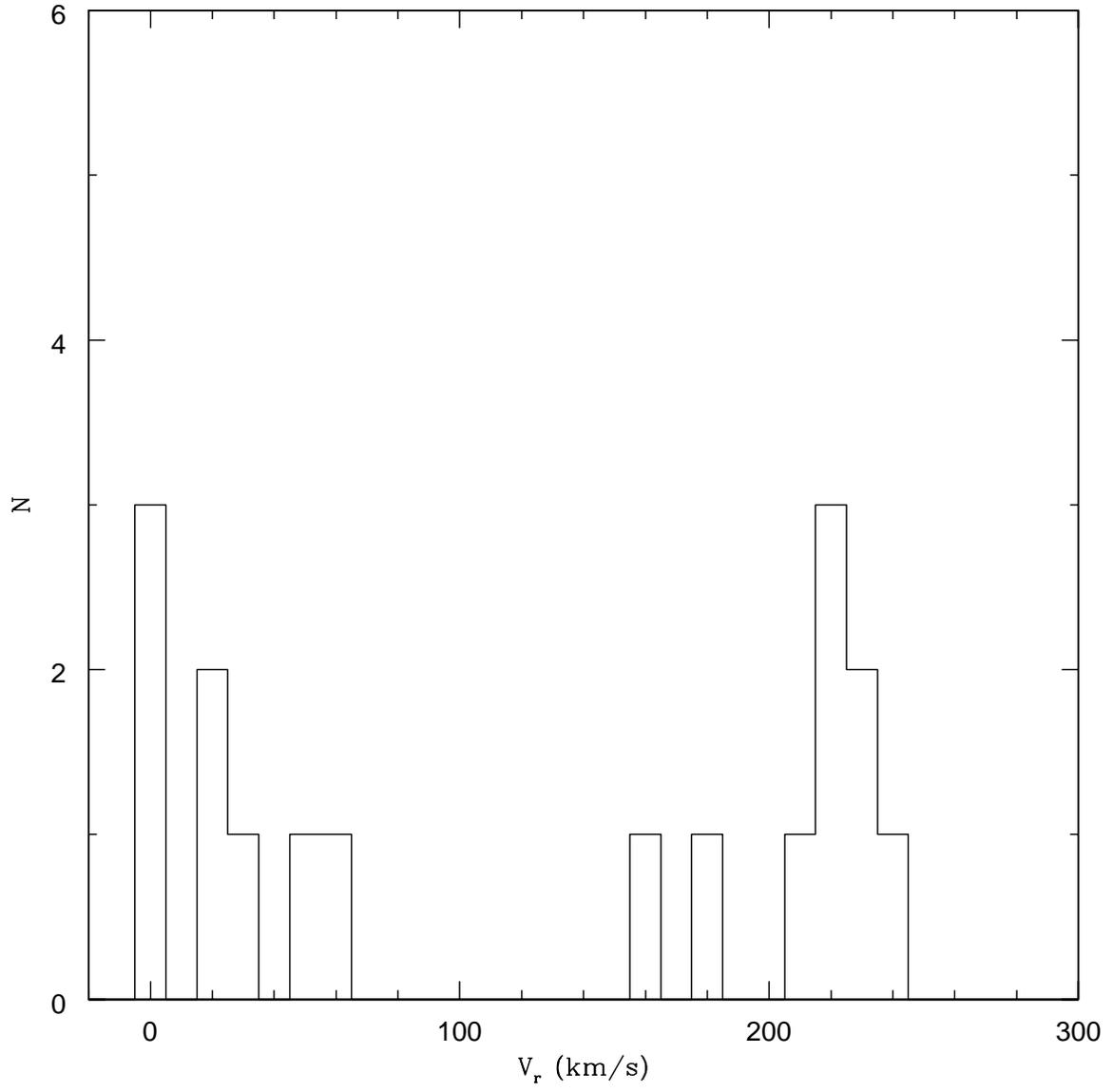}
\caption{RV distribution for blue stars with $M-T_2 < 0.75$.}
\end{figure}

\begin{figure}
\epsscale{0.57} 
\plotone{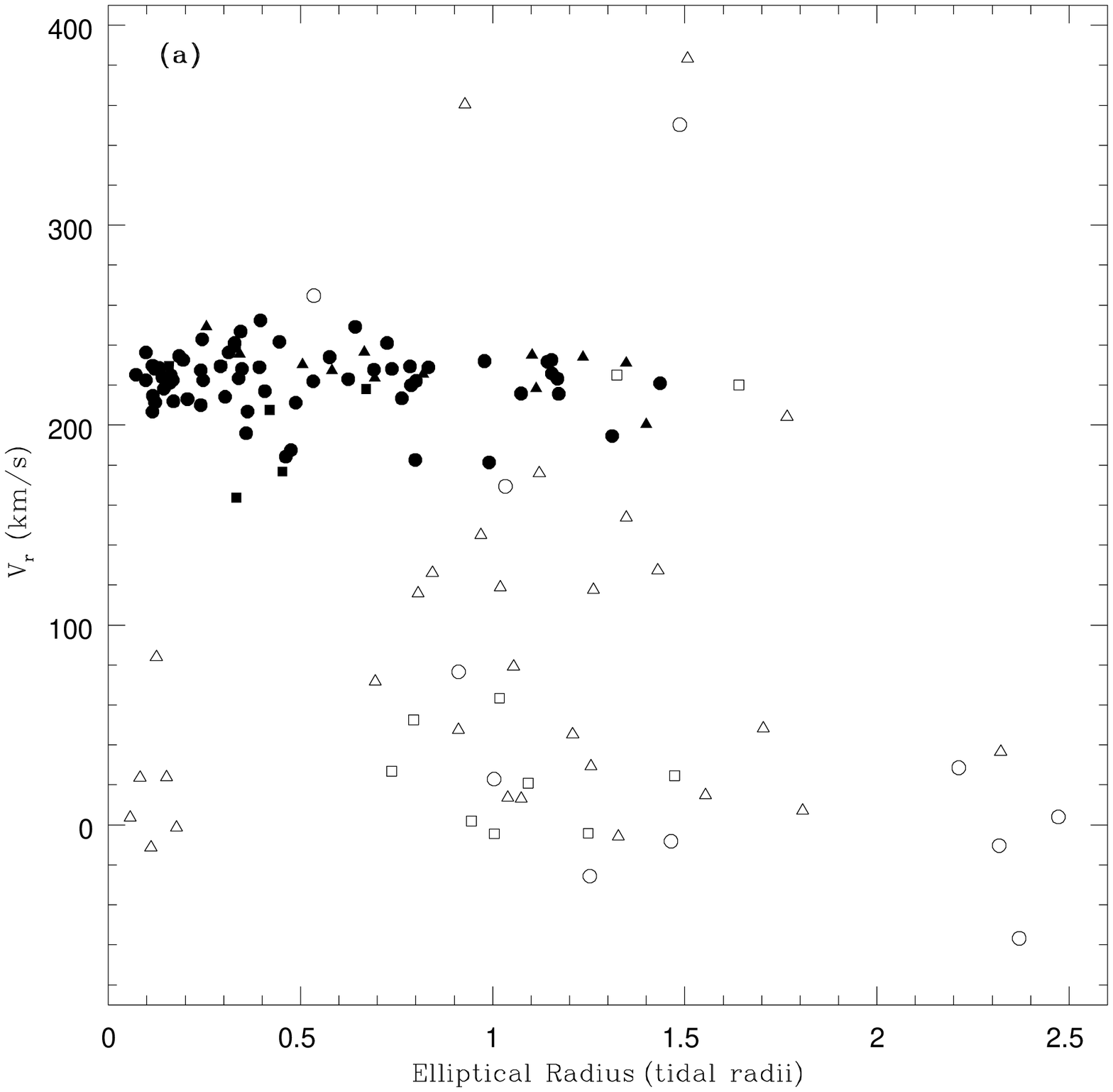} 
\plotone{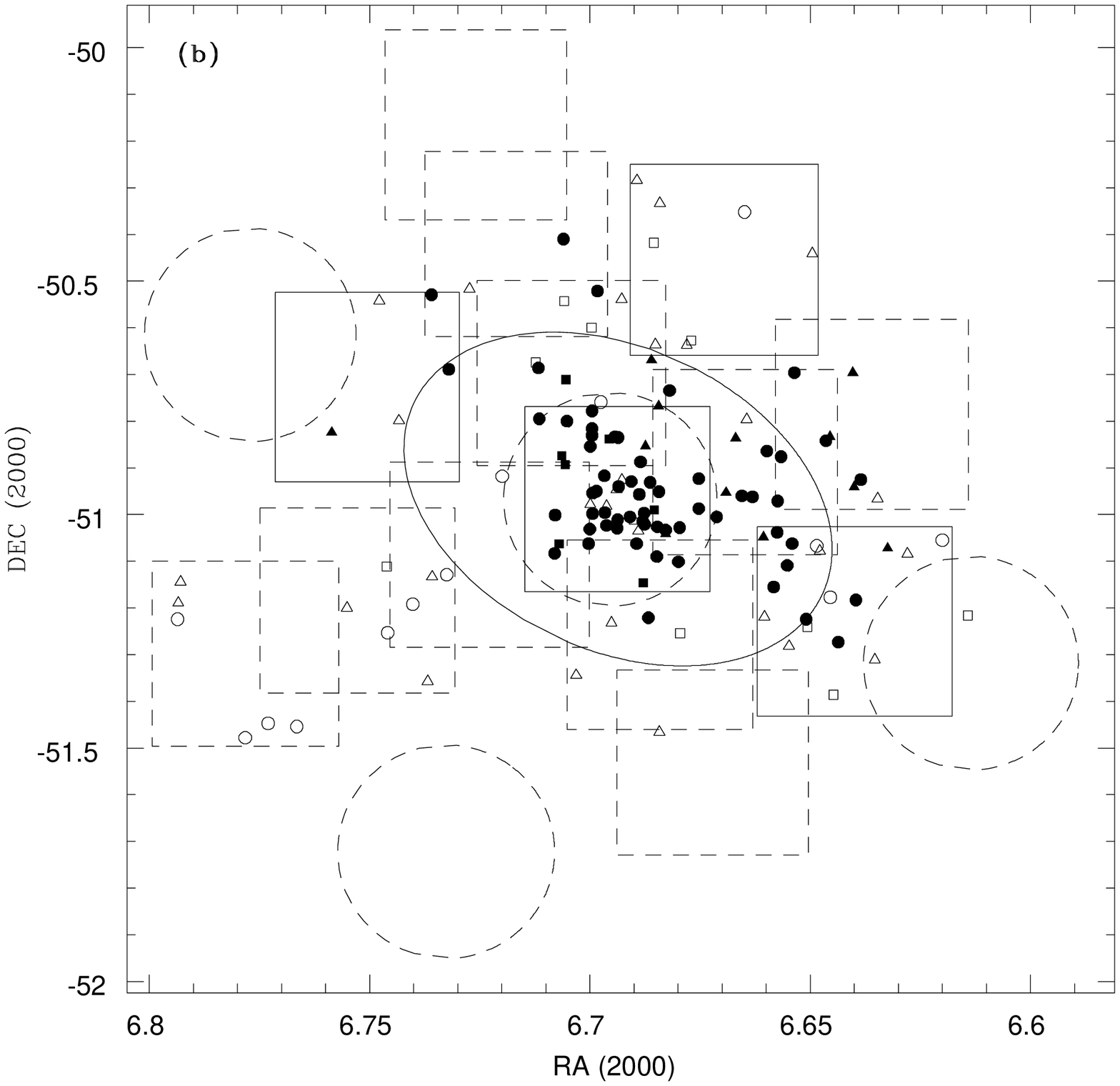}
\caption{(a) Radial velocities for all stars from Table 5
  as a function of their elliptical distance from the Carina center.
  Stars photometrically selected as Carina giants are shown as circles,
  blue stars are shown as squares, and all other stars are shown as
  triangles.  (b) Spatial distribution of Table 5 stars.  The solid
  ellipse is the \citeauthor{IH95} tidal radius.  As in
  \citeauthor{paperII}, the large squares show the Swope telescope CCD
  frame positions and the large circles show the du Pont pointings; when
  drawn with solid lines the data were taken in photometric conditions.
  In both panels, stars considered to be Carina members (and class
  ``Y?'') are shown with filled symbols.}
\end{figure}

\end{document}